\documentclass[aps,prx,reprint,superscriptaddress,longbibliography]{revtex4-2}
\PassOptionsToPackage{dvipsnames,svgnames,table}{xcolor}
\usepackage{bm}
\usepackage{placeins}
\usepackage{graphicx}
\usepackage{amsmath}
\usepackage{amsfonts}
\usepackage{amssymb}
\usepackage{color}
\usepackage{float}
\usepackage[caption=false]{subfig}
\usepackage{xcolor}
\usepackage{soul}
\usepackage{times}
\usepackage{hyperref}
\hypersetup{
  pdfstartview={FitH},
  breaklinks=true,
  colorlinks=true,
  linkcolor=NavyBlue,  % internal links (sections, figures, equations)
  citecolor=Maroon,    % bibliography links
  filecolor=NavyBlue,
  urlcolor=NavyBlue
}

\newcommand{\Tr}{\operatorname{Tr}}
\newcommand{\Real}{\mathrm{Re}}
\newcommand{\Imag}{\mathrm{Im}}
\newcommand{\Svir}{\sigma_{\mathrm{vir}}}
\newcommand{\Avir}{A_{\mathrm{vir}}}
\newcommand{\kk}{\mathbf{k}}
\newcommand{\qq}{\mathbf{q}}
\newcommand{\Cinf}{C_{\infty}}
\graphicspath{{}}

\begin{document}

\title{Symmetry selection rule for the band-edge shift current in two dimensions}

\author{Felipe P\'erez Riffo}
\affiliation{Departamento de F\'isica, Universidad T\'ecnica Federico Santa Mar\'ia,
Casilla 110-V, Valpara\'iso, Chile}
\author{Mat\'ias Castro Schnaidt}
\affiliation{Departamento de F\'isica, Universidad T\'ecnica Federico Santa Mar\'ia,
Casilla 110-V, Valpara\'iso, Chile}
\author{Leonor Chico}
\affiliation{GISC, Departamento de F\'isica de Materiales, Facultad de Ciencias F\'isicas, Universidad Complutense de Madrid, Plaza de las Ciencias 1, 28040 Madrid, Spain}
\author{L.~E.~F.~Foa~Torres}
\affiliation{Departamento de F\'isica, Facultad de Ciencias F\'isicas y Matem\'aticas,
Universidad de Chile, Santiago 837.0415, Chile}
\author{Eric Su\'arez Morell}
\email[Corresponding author: ]{eric.suarez@usm.cl}
\affiliation{Departamento de F\'isica, Universidad T\'ecnica Federico Santa Mar\'ia,
Casilla 110-V, Valpara\'iso, Chile}
\date{\today}

\begin{abstract}
The shift current is the intrinsic bulk photovoltaic response of a crystal without an
inversion center. In graphene multilayers, calculations report large band-edge shift
currents that reverse sign under a gate, a behavior that neither the quantum metric nor
the Berry curvature captures. We show that this follows from a symmetry principle. At
the absorption edge of a gapped, inversion-broken two-dimensional Dirac-like system, an
emergent rotational symmetry forbids the response, and the trigonal warping switches it
on linearly in its strength, an angular selection rule that unifies multilayer graphene
and the kagome lattice. The released current is governed by a signed, detuning-weighted
three-point Bargmann invariant: the optical amplitude stays fixed while the sign
alignment of the Bargmann triangles grows with the warping, invisible to any
positive-definite figure of merit. In bilayer and trilayer graphene the gate alone
reverses the sign, making the band-edge shift current a gate-switchable bulk
photovoltaic response.
\end{abstract}

\maketitle

% #####################################################################
\section{Introduction}
\label{sec:intro}
% #####################################################################
The bulk photovoltaic effect is the direct current that a homogeneous, inversion-broken crystal generates under uniform illumination, without the $p$--$n$ junction of a conventional solar cell~\cite{vonBaltz1981,Sipe2000}. A built-in field does not limit the effect, so the photovoltage can exceed the band gap. This mechanism
enables junction-free energy conversion and
photodetection~\cite{OrensteinMoore2021,DaiRappe2023}. Under linearly polarized light, the
dominant contribution is the \emph{shift current} (SC): the coherent real-space displacement
of the electronic wavepacket during an interband optical
transition~\cite{Sipe2000,Morimoto2016}. Ferroelectric engineering of the SC has recently
driven two-dimensional bulk-photovoltaic devices to record short-circuit currents and higher
conversion efficiency~\cite{Feng2025}. First-principles calculations reproduce the measured
photocurrents of ferroelectrics and have led to design rules for large
responses~\cite{YoungRappe2012,Cook2017,TanRappe2019,FregosoMorimotoMoore2017,RangelFregoso2017}.

The displacement behind the SC has a geometric origin: each transition shifts the
electron's charge center by the \emph{shift vector}, a difference of Berry connections
between the two bands involved~\cite{Morimoto2016,AhnGuoNagaosa2020}. The SC is therefore
a quantum-geometric property of the Bloch states. Resonant optical responses are governed by the geometry of the Bloch states, and the
geometric objects are organized by the number of bands they involve. The simplest is the
interband dipole, the optical transition matrix element between two bands. Products of two
dipoles are gauge invariant and give the two-band objects: the quantum metric tensor (QMT)
and the Berry curvature. Processes that connect three or more bands require higher
invariants~\cite{AhnGuoNagaosaVishwanath2022,JiangHolderYan2025}.

The same hierarchy underlies responses from the nonlinear Hall
effect~\cite{SodemannFu2015} to the quantized circular photogalvanic
effect~\cite{deJuan2017}. Beyond the two-band objects, the SC involves a \emph{multistate}
(Bargmann) geometry: closed cycles of interband
dipoles~\cite{AvdoshkinPopov2023,Multistate2024,ProjectorCalculus2025}. In graphene
multilayers, this multiband channel strongly enhances the response~\cite{Chen2024}.

SC-like photoresponses also arise outside the closed-crystal, weak-field regime treated
here, for instance as nonadiabatic charge pumping in finite, contacted devices, where asymmetry of the
device itself, rather than bulk noncentrosymmetry, can supply the required symmetry breaking~\cite{Bajpai2019}.
The present work instead isolates the bulk, second-order-in-field mechanism and its symmetry protection near
band-edge touchings.

Two-dimensional Dirac-like systems are the cleanest arena in which to ask what controls this geometry at the absorption edge. Gapped graphene multilayers realize a ladder of chiral band touchings whose pseudospin winds
$n$ times with the number of layers: linear ($n{=}1$), quadratic ($n{=}2$), and cubic
($n{=}3$)~\cite{McCannKoshino2013,KoshinoMcCann2009}, all sharing the lattice's trigonal ($C_3$) warping. The optical response of these structures is dictated by the stacking geometry: for instance, a relative twist alone renders the bilayer chiral and circularly dichroic~\cite{SuarezMorellCD2017}, and rotating a single layer of the trilayer reshapes its absorption spectrum across the infrared and visible~\cite{Correa2014}.

A series of recent calculations reports large band-edge SCs on this platform that reverse
sign under a gate or displacement
field~\cite{Zheng2023,ABCtrilayer2024,GiantSC2025,CentrosymBPVE2025}. Neither observation
is explained. The sign escapes the QMT and the Berry curvature~\cite{ABCtrilayer2024}, and
the magnitude has not been traced to any coupling; in particular, the influence of the
trigonal warping has never been studied in the multilayers.

In a gapped Dirac cone, including the trigonal warping switches on the
SC~\cite{KimMorimoto2017}, and the two-dimensional band-edge response diverges as the gap
closes~\cite{YoshidaMurakami2025}. The question extends beyond graphene. Inversion-broken
kagome metals show an SC tied to a flat band~\cite{Liu2025kagomeSC}, and the problem
thereby joins the flat-band quantum geometry of correlated two-dimensional systems, a
program that traces back to the flat bands of slightly twisted bilayer
graphene~\cite{SuarezMorell2010,TormaPeottaBernevig2022}. What is missing is a single
mechanism that sets both the magnitude and the sign that the gate reverses.

In this work, we show that at any gapped, winding-$n$ Dirac-like touching in two
dimensions, an emergent continuous rotational symmetry forbids the band-edge SC. The
prohibition is intrinsically two-dimensional. Three interband dipoles cannot build a
rotational scalar under the planar rotation group SO(2), whereas under SO(3) they can
(Sec.~\ref{sec:vanish}).

Trigonal warping reduces the emergent symmetry to the crystal's $C_3$ and supplies the
missing scalar. The current then switches on linearly in the warping strength, through a
lattice coupling of gauge-invariant angular charge $\pm3$. The released quantity is a
signed, detuning-weighted three-point Bargmann invariant, built on the three states of the
virtual transition. Its growth is a phase-coherence effect; the optical amplitude stays
fixed while the sign alignment of the Bargmann triangles grows with the warping. We define
a bounded coherence factor, the ratio of the net signed weight to its unsigned magnitude,
that captures this alignment. It carries the linear growth of the band edge and its
gate-driven, parameter-free sign reversal in bilayer and trilayer; the remainder is a
smooth direct-channel background. This sign is what metric-based figures of merit miss.
The QMT fails at the band edge; the signed invariant reverses where the metric only grows.
The rule unifies monolayer, bilayer, and trilayer graphene with an inversion-broken kagome
lattice, where the flat band realizes the channel intrinsically, without warping.

The paper is organized as follows. Section~\ref{sec:framework} sets up the SC as multistate quantum
geometry and defines the three-point invariant and its coherence factor. Section~\ref{sec:selection} derives
the angular selection rule and its two-dimensional character.  Sections~\ref{sec:family} and~\ref{sec:kagome} develop the graphene family, with the bilayer's microscopic anatomy, and the kagome lattice complement. Section~\ref{sec:closedform} solves the bilayer
release analytically. Section~\ref{sec:metric} assesses
the QMT as a band-edge figure of merit, and Sec.~\ref{sec:experiment} outlines an experimental
proposal, before a concluding discussion (Sec.~\ref{sec:discussion}).

% #####################################################################
\section{Shift current as multistate quantum geometry}
\label{sec:framework}
% #####################################################################

\subsection{The shift current: transition dipole and shift vector}
\label{sec:sipe}
Under uniform illumination linearly polarized along the $b$ direction, the DC SC is
$j^a=\sigma^{a;bb}(\omega)\,|E^b(\omega)|^2$. For a clean band insulator at $T=0$ K the shift conductivity
takes the intuitive Sipe--Shkrebtii form~\cite{Sipe2000,Morimoto2016,Cook2017}
\begin{equation}
\sigma^{a;bb}(\omega)=-\frac{\pi e^3}{\hbar^2}\!\int\!\frac{d^2k}{(2\pi)^2}
\sum_{n,m} f_{nm}\!
\underbrace{|r^b_{nm}|^2}_{\text{rate}}\!
\underbrace{R^{a;b}_{nm}}_{\substack{\text{shift}\\\text{vector}}}\!
\underbrace{\delta(\omega_{nm}-\omega)}_{\text{resonance}},
\label{eq:scshift}
\end{equation}
 where the meaning of the factors in the integrand is indicated therein. The occupation difference $f_{nm}=f_n-f_m$ restricts the sum
to transitions from filled to empty bands; the delta function enforces a vertical
 transition at $\omega_{nm}=\varepsilon_n-\varepsilon_m$; and the transition is
weighted by two geometric quantities. The first is the interband transition dipole
\begin{equation}
r^a_{nm}=\frac{v^a_{nm}}{i\,\omega_{nm}}\quad(n\neq m),\qquad
v^a_{nm}=\langle u_n|\partial_{k_a}H|u_m\rangle,
\label{eq:dipoledef}
\end{equation}
whose squared modulus $|r^b_{nm}|^2$ is the optical transition probability. The second is the \emph{shift vector}
\begin{equation}
R^{a;b}_{nm}=\partial_{k_a}\arg r^b_{nm}-\big(A^a_{nn}-A^a_{mm}\big),
\label{eq:shiftvector}
\end{equation}
with $A^a_{nn}=\langle u_n|i\partial_{k_a}u_n\rangle$ the intraband Berry connection: the shift vector is the difference between
the intracell positions (charge centers) of the electron in the two bands, a
real-space displacement of the wavepacket during the vertical transition~\cite{Morimoto2016,AhnGuoNagaosa2020}.
Being a coordinate difference, $R^{a;b}_{nm}$ is gauge invariant and \emph{odd under inversion}, which is why a
nonzero SC requires a crystal without an inversion center. This is a statement about the
closed, translationally-invariant periodic crystal treated by the formalism of Eq.~\eqref{eq:scshift}; in
finite, contacted devices the required asymmetry can instead be supplied by the leads rather than the bulk
crystal, permitting nonadiabatic SC-like photocurrents even from a centrosymmetric bulk
material~\cite{Bajpai2019}.

\begin{figure}[t]
\centering
\includegraphics[width=\linewidth]{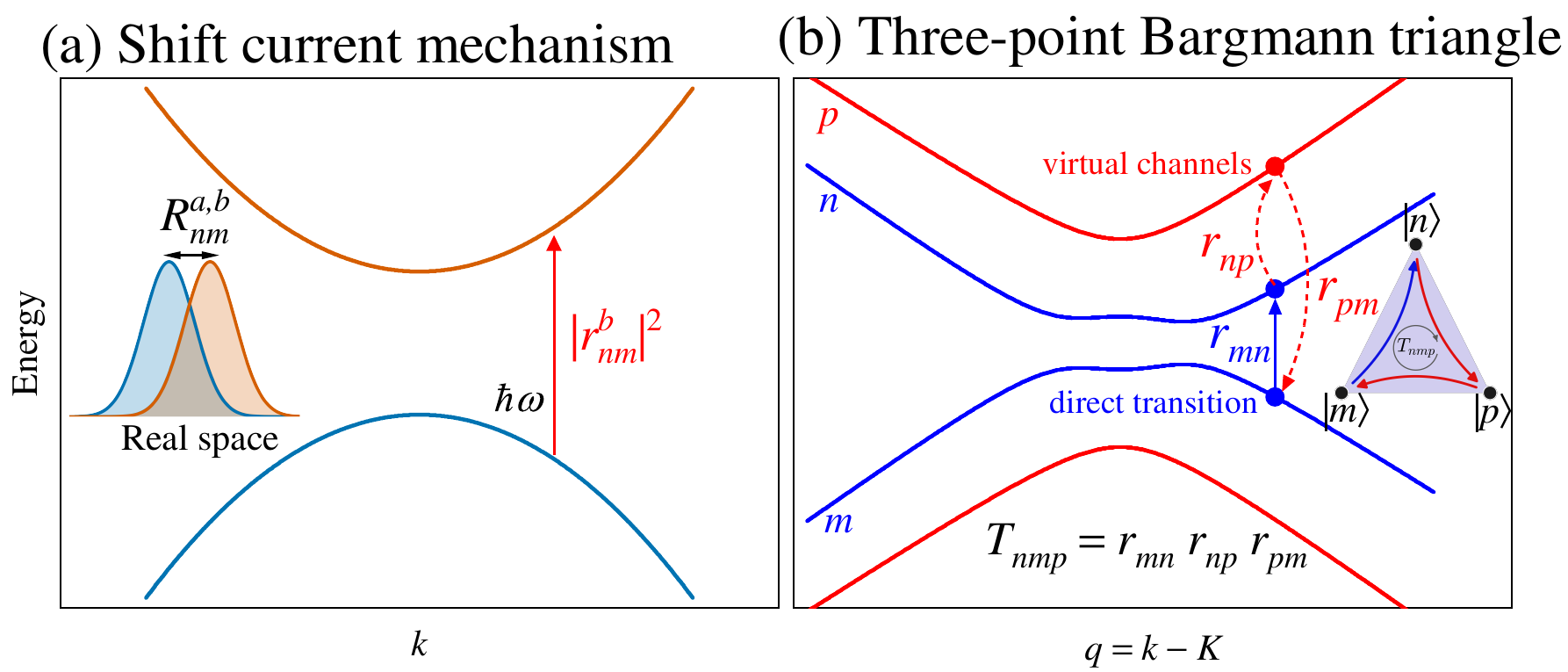}
\caption{(a) SC mechanism: a vertical interband transition promotes
an electron at fixed $\kk$ [rate $|r^b_{nm}|^2$], displacing its real-space charge center by the shift vector
$R^{a;b}_{nm}$; the photocurrent is the coherent sum of these shifts.
(b) Gapped AB-bilayer bands near $K$ (low-energy optical pair $m,n$ in blue; dimer bands in red): the
resonant pair and a dimer spectator $p$ form the three-point Bargmann triangle $m\!\to\!n\!\to\!p\!\to\!m$
with interband connections $r_{nm}$ (optical) and $r_{np},r_{pm}$ (virtual).}
\label{fig:mechanism}
\end{figure}

Equation~\eqref{eq:scshift} therefore reads as a product of a \emph{rate} and a \emph{displacement}. Electrons
are promoted vertically at a rate
$\propto|r^b_{nm}|^2$, each promotion shifts the charge center by $R^{a;b}_{nm}$, and the photocurrent is the
coherent sum of these microscopic shifts [Fig.~\ref{fig:mechanism}(a)].

\subsection{Direct and virtual channels}
\label{sec:split}
To expose the multiband structure that will rule the band edge, it is convenient to express the shift
vector as a covariant momentum derivative. Writing the diagonal integrand as
$I^{a;bb}_{nm}=|r^b_{nm}|^2R^{a;b}_{nm}=\Imag[\,\bar r^{\,b}_{nm}\,r^{b}_{nm;a}\,]$, the conductivity of
Eq.~\eqref{eq:scshift} is reproduced by the operational form~\cite{Sipe2000,AhnGuoNagaosaVishwanath2022}
\begin{equation}
\sigma^{a;bc}(\omega)=-\frac{\pi e^3}{\hbar^2}\!\int\!\frac{d^2k}{(2\pi)^2}
\sum_{n,m} f_{nm}\,I^{a;bc}_{nm}\,\delta(\omega_{nm}-\omega),
\label{eq:sipe}
\end{equation}
with interband integrand
\begin{equation}
I^{a;bc}_{nm}=\Imag\!\big[\,r^b_{mn}\,r^{c}_{nm;a}+r^c_{mn}\,r^{b}_{nm;a}\,\big],
\label{eq:integrand}
\end{equation}
built from the same dipole~\eqref{eq:dipoledef} and the generalized (covariant) derivative
\begin{equation}
r^{b}_{nm;a}=\big[\partial_{k_a}-i(A^a_{nn}-A^a_{mm})\big]\,r^b_{nm},
\label{eq:gend}
\end{equation}
whose $U(1)$-covariant construction is what makes each $I^{a;bc}_{nm}$ gauge invariant. Carrying out
$\partial_{k_a}$ with the velocity sum rule splits the generalized derivative into two separately
gauge-invariant groups. The \emph{direct} part, involving only the resonant pair $(m,n)$,
\begin{equation}
r^{a}_{nm;b}\big|_{\rm dir}=\frac{i}{\omega_{nm}}\!\left[
\frac{v^a_{nm}\Delta^b_{nm}+v^b_{nm}\Delta^a_{nm}}{\omega_{nm}}-W^{ab}_{nm}\right],
\label{eq:direct}
\end{equation}
 
where $\Delta^a_{nm}=v^a_{nn}-v^a_{mm}$ and $W^{ab}_{nm}=\langle n|\partial_{k_a}\partial_{k_b}H|m\rangle$,
reproduces the familiar two-band shift-vector physics, while the \emph{virtual} part, summed over all
spectator bands $p\neq n,m$,
\begin{equation}
r^{a}_{nm;b}\big|_{\rm vir}=\frac{i}{\omega_{nm}}\sum_{p\neq n,m}
\Big(\frac{v^a_{np}\,v^b_{pm}}{\omega_{pm}}-\frac{v^b_{np}\,v^a_{pm}}{\omega_{np}}\Big),
\label{eq:virt}
\end{equation}
allows the electron to make a virtual excursion through a third band; it is an irreducibly multiband process, absent in any
two-level description. For the $C_3$- and mirror-symmetric systems studied here the third-rank tensor has a
single independent component, $\sigma^{yyy}=-\sigma^{yxx}$ with $\sigma^{xxx}=0$~\cite{PeronaciOka2021}. We
therefore work throughout with the diagonal $\sigma^{yyy}$. This constrain makes the virtual channel collapses to one
transparent object, as we now show.

\subsection{The three-point Bargmann invariant and the coherence factor}
\label{sec:bargmann}
Individual overlaps of quantum states carry arbitrary phases; the gauge-invariant
content lives in products of overlaps taken around a closed
cycle~\cite{Bargmann1964,Pancharatnam1956,Mukunda1993}.
The simplest such invariant beyond a single pair is the three-point (triangle) Bargmann invariant, and it is
the very object the virtual channel of Eq.~\eqref{eq:virt} constitutes. We rewrite the velocities as connections,
$v^a_{ij}=i\,\omega_{ij}\,r^a_{ij}$, substitute into Eq.~\eqref{eq:virt}, and contract with
$\bar r^{\,b}_{nm}$. For the diagonal component ($a{=}b{=}y$), the two terms of
Eq.~\eqref{eq:virt} share the factor $v^y_{np}v^y_{pm}$ and merge into a single triple
product per spectator. The energy denominators collapse to a single dimensionless weight,
$\omega_{pm}-\omega_{np}=-D_p\,\omega_{nm}$, and the diagonal virtual weight becomes
\begin{equation}
\Svir=\sum_{p\neq n,m}D_p\,\Real\,T_{nmp},\qquad T_{nmp}=r_{mn}\,r_{np}\,r_{pm},
\label{eq:closed}
\end{equation}
Each $T_{nmp}$ is a closed loop $m\to n\to p\to m$ in band space
[Fig.~\ref{fig:mechanism}(b)]. The weight is the detuning
$D_p=(\varepsilon_n+\varepsilon_m-2\varepsilon_p)/(\varepsilon_n-\varepsilon_m)$, the
energy offset of the spectator from the midpoint of the optical pair in units of the pair
splitting, negative when $p$ lies above it.

Two properties make $T_{nmp}$ the relevant object. First, it is gauge invariant;
every band index appears once as a bra and once as a ket. It is not the three-projector
invariant $\Tr[P_nP_mP_p]$ of the eigenstates at a single $\kk$, which vanishes because
distinct eigenstates are orthogonal; it is built from the momentum-space derivatives that
couple the bands, and it encodes how the states vary across the Brillouin zone
(BZ)~\cite{AhnGuoNagaosaVishwanath2022,Multistate2024}. Second, it is \emph{signed}, a
consequence of involving three states. A product of two dipoles gives the positive
$|r_{nm}|^2$; a closed cycle through three states retains a gauge-invariant phase, so
$\Real\,T_{nmp}$ can take either sign. How this sign enters the response is set jointly
with the detuning weight, as we show next.

Inherited from the energy denominators of the virtual process, the detuning weight is
sign-indefinite; spectators above and below the pair midpoint enter with opposite signs
and bear directly on the SC. The arrangement of the levels thereby co-determines the net
sign, together with the wavefunction geometry. Geometry alone can even mislead; in the
bilayer, the unweighted sum $\sum_p\Real\,T_{nmp}$ anti-correlates with the response.
$\Svir$ is the physical combination, the Bargmann geometry of the triangles weighted by
the spectrum.

The two-point companion of Eq.~\eqref{eq:closed} is the quantum geometric tensor
$Q^{ab}_n=\sum_{m\neq n}r^a_{nm}r^b_{mn}$. Its real part is the QMT, a positive sum of
squares that lower-bounds the Wannier spread~\cite{Marzari1997}. In the language of the modern theory of polarization, the metric is an even second
moment (a variance), while the three-point invariant is an odd third moment (a skewness).
The metric measures how localized the states are; it carries no information about the displacement of the wavepacket. 

Several of the phenomena studied below are governed by the collective sign of the three-point invariant
rather than by its magnitude. It is therefore convenient to define a single bounded quantity, the
\emph{coherence factor}, as the ratio of the net signed weight to its positive, incoherent companion,
\begin{equation}
C=\frac{\sum_{\kk,p} D_p\,\Real\,T_{nmp}}
       {\sum_{\kk,p} \big|D_p\,\Real\,T_{nmp}\big|}\in[-1,1].
\label{eq:coherence}
\end{equation}
The numerator is the BZ integral of $\Svir$ [Eq.~\eqref{eq:closed}], and the
denominator $\Avir\equiv\sum_{\kk,p}|D_p\,\Real\,T_{nmp}|$ is the total multiband
amplitude irrespective of sign; thus $C=\pm1$ when all triangles share one sign and
$C\to0$ under cancellation.

The coherence factor
separates magnitude from phase alignment. It is this separation, rather than any change in optical amplitude, that
governs the results below. 

The coherence invoked here is a static property of the momentum-space
virtual quantum connection, integrated over a resonance shell at fixed $\hbar\omega$, distinct from the
dynamical, real-space coherence (off-diagonal elements of the nonequilibrium density matrix $\rho^{\rm
neq}(t)$) shown to carry SC signals over macroscopic distances in driven, finite quantum-transport
devices~\cite{Bajpai2019}. Whether these two notions of coherence are related is not addressed here.

Our main results are the angular selection rule that determines when $\Svir$ is
forbidden and how it is released; its bounded coherence factor $C$ as the band-edge figure
of merit; and the gate-tunable sign reversal that follows as a clear-cut experimental
signature. All three are demonstrated for, and claimed only within, one class of
Hamiltonians, gapped Dirac-like and flat-band touchings in two dimensions with a $C_3$
little group (Sec.~\ref{sec:discussion}).

% #####################################################################
\section{The angular selection rule}
\label{sec:selection}
% #####################################################################

\subsection{Emergent rotational symmetry of a gapped winding-\texorpdfstring{$n$}{n} touching}
\label{sec:emergent}
A two-band Hamiltonian can always be written as $H=\mathbf d(\qq)\cdot\bm\sigma$ (up to a particle-hole term
$\propto\sigma_0$ that we omit). The Dirac-like systems we study are
defined by the winding of the in-plane pseudospin. Near the touching, the $K$ point, the
leading order of the $k\cdot p$ expansion reduces the in-plane part to a single isotropic
winding, $d_x+id_y=|\mathbf d_\parallel(q)|\,e^{in\theta}$, and a diagonal (sublattice)
mass $d_z$ opens the gap. For the graphene family, the lattice sets the winding equal to
the number of layers; $n{=}1$ for the monolayer (linear Dirac), $n{=}2$ for AB
(quadratic), and $n{=}3$ for ABC (cubic)~\cite{McCannKoshino2013,KoshinoMcCann2009}.
Equivalently, each component of the Bloch eigenvector carries a definite phase
$e^{i\ell_j\theta}$, set by a sublattice charge $\ell_j$. In the two-band form, the two
charges differ by the winding $n$ itself; in the full linearized multilayers, the same
holds for the two non-dimer ends (explicit Hamiltonians and charges in
App.~\ref{app:models}).

The gap opens through a term diagonal in the sublattice basis, a Semenoff mass in the
monolayer~\cite{Semenoff1984} or the interlayer bias $\delta$ in the bilayer and
rhombohedral trilayer. Such a term preserves an \emph{emergent} continuous rotation,
because it commutes with the diagonal phase rotation $U(\varphi)$ below. Writing the
momentum measured from the $K$ point as $\qq=q(\cos\theta,\sin\theta)$ and letting
$R_\varphi$ rotate it by $\varphi$, the gapped Hamiltonian obeys
\begin{equation}
H_0(R_\varphi\qq)=U(\varphi)\,H_0(\qq)\,U^\dagger(\varphi),\quad U(\varphi)=\mathrm{diag}(e^{i\ell_j\varphi}):
\label{eq:emergent}
\end{equation}
rotating the momentum by $\varphi$ is undone by a sublattice phase rotation of charge
$\ell_j$. Equivalently, $H_0$ commutes with the total angular momentum
$J=-i\partial_\theta-\hat L$, the sum of an orbital part and a sublattice (pseudospin)
part, with $\hat L=\mathrm{diag}(\ell_j)$; the charge $\ell_j$ is thus the intrinsic
angular momentum that site $j$ contributes to $J$. This emergent continuous symmetry
$\Cinf$ is larger than the crystal's $C_3$ rotational symmetry. Two consequences follow.
The energies depend only on $|q|$, not on $\theta$. The eigenvectors are $J$-eigenstates,
obtained from their $\theta{=}0$ form by $U(\theta)$, which is how each component acquires
its phase $e^{i\ell_j\theta}$.

\subsection{Vanishing of the band edge and its two-dimensional origin}
\label{sec:vanish}
The shift conductivity is a rank-three tensor, and the band-edge response is its rotational scalar, the
angular average ($\ell{=}0$) of the integrand of Eq.~\eqref{eq:sipe}. The emergent $\Cinf$ symmetry forbids the band-edge response. The linearized-model argument establishing this is exact and proceeds in four steps. (i) At zero warping, the only $\kk$-dependent matrix elements of $H_0$ are the linear intralayer hoppings, whose derivative is constant, so the velocity matrix $\partial_{k_a}H_0$ has $\theta$-independent entries. (ii) With a constant velocity matrix and eigenvectors
carrying $e^{i\ell_j\theta}$, every interband dipole winds as $r^a_{ij}\sim e^{\pm i\theta}$; it is a vector,
of charge $\ell{=}\pm1$ under the emergent $\Cinf$. (iii) The band-edge object (the two-band integrand in the monolayer,
the triple product $r_{mn}r_{np}r_{pm}$ in the multiband systems) is a product of three such factors, so its
total winding is $(\pm1)+(\pm1)+(\pm1)\in\{\pm1,\pm3\}$, always odd and never $0$. (iv) The angular average
therefore vanishes,
\begin{equation}
I_0(q)\equiv\frac1{2\pi}\!\int_0^{2\pi}\!\!d\theta\,I(q,\theta)=0\qquad(\text{no }\ell{=}0\text{ at }\gamma_3{=}0),
\label{eq:vanish}
\end{equation}
so the band-edge SC is forbidden even though the mass $\delta$ already breaks inversion.
The obstruction follows from transformation properties alone; no value of a matrix
element can evade it. $\sigma^{yyy}$ is a rotational scalar built from three vectors, and
under a continuous planar rotation three vectors cannot sum to a scalar.

The prohibition is specific to two dimensions, for two independent reasons. First, the SC is
inversion-odd, and in 2D spatial inversion \emph{is} the $C_2$ rotation. The even orders $C_2,C_4,C_6$ all
contain it and forbid the response outright, so a nonzero band edge requires both the broken inversion that the mass supplies and a lattice anisotropy able to reach $\ell{=}0$. Second, the scalar-counting is a statement about the planar rotation group $SO(2)$ (the
emergent $\Cinf$ in group-theoretic language), whose irreducible representations are the
integers; three vectors ($\ell{=}\pm1$) can only build $\ell\in\{\pm1,\pm3\}$. Under $SO(3)$ the same three vectors \emph{do} contain a scalar, the triple product
$\varepsilon_{abc}$, so an emergent rotational symmetry does not forbid the band-edge SC at a
three-dimensional Weyl touching\footnote{The distinction is the rank of the invariant antisymmetric tensor. In
three dimensions the Levi-Civita symbol $\varepsilon_{abc}$ has rank three, so three vectors contract to a
rotational scalar. In two dimensions the invariant tensor $\varepsilon_{ab}$ has rank two: it pairs only
\emph{two} vectors into a scalar, and any product of an odd number of vectors carries odd angular momentum,
never $\ell{=}0$.}. It is in this precise sense a two-dimensional selection rule.

\subsection{The angular-charge rule and linear release}
\label{sec:weight}
The forbidden band-edge response can be restored in two ways. Keeping $\Cinf$ intact requires a non-vertical
transition; at finite momentum transfer a photon-drag current is symmetry-allowed~\cite{RofiiHasdeo2026}. We take the standard route, keeping the vertical ($q{=}0$) transition and breaking $\Cinf$ with the crystalline warping. A real
lattice retains only $C_3$, which conserves angular momentum modulo $3$; of the base harmonics $\{\pm1,\pm3\}$
only $\ell{=}\pm3\equiv0$ becomes a $C_3$ scalar, and it is the unique channel that can feed the band edge.

Which coupling supplies that channel is not decided by the orbital winding of its
momentum factor alone. Under the combined rotation-and-gauge operation of
Eq.~\eqref{eq:emergent}, a matrix element $(H_1)_{ab}$ of the warping term acquires a
phase from two sources, the orbital winding of its momentum factor and the sublattice
charges of the two sites it connects. The net phase defines the gauge-invariant angular
charge of the coupling,
\begin{equation}
\big[U(\varphi)\,H_1(R_{-\varphi}\qq)\,U^\dagger(\varphi)\big]_{ab}=e^{iw\varphi}\,(H_1)_{ab},
\label{eq:weight}
\end{equation}
with $w=(\ell_a-\ell_b)-w_{\rm hop}$, the angular momentum the coupling transfers.
Two couplings with the same momentum winding can therefore carry different charges when
they connect different sites; the bilayer's $\gamma_3$ and $\gamma_4$, both linear in
momentum, carry $w=-3$ and $w=0$ (Apps.~\ref{app:lcount} and~\ref{app:numerics}). The
releasing coupling is the one with $w{=}\pm3$; it connects the sublattices that the
touching leaves disconnected, and it transfers three units of angular momentum even when
its momentum factor winds by only $1$ or $2$. For each member of the graphene family below, the
physical warping term carries this very charge (App.~\ref{app:lcount}).

The charge also fixes the order in $\gamma_3$ at which the response turns on. A single action of the $w{=}\pm3$ coupling shifts every base harmonic by
$\pm3$,
\begin{equation}
\ell{=}\pm3\ \longrightarrow\ \{0,\pm6\},\qquad \ell{=}\pm1\ \longrightarrow\ \{\mp2,\pm4\},
\label{eq:shift}
\end{equation}
so the base $\ell{=}\pm3$ reaches the scalar $\ell{=}0$ ($3-3=0$) while $\ell{=}\pm1$ lands on $\pm2,\pm4$, which never reach $\ell{=}0$. Because $\ell{=}0$ is forbidden at zeroth order [Eq.~\eqref{eq:vanish}] and first reached by a single
action of the warping term, the band-edge response turns on at \emph{first order},
$\Svir=B\,\gamma_3+O(\gamma_3^3)$, identically for AB and ABC. 
%The cubic touching of ABC enters the windings, not the release order.

The response is in fact odd in $\gamma_3$,
$\sigma^{yyy}(-\gamma_3)=-\sigma^{yyy}(\gamma_3)$, so the expansion contains no even
powers and the linear law is corrected only at $O(\gamma_3^{3})$. This strict oddness is enforced by the vertical mirror $M_x$, the $C_{3v}$ element
that also sets $\sigma^{xxx}=0$; the mirror maps the lattice with $-\gamma_3$ onto the
lattice with $+\gamma_3$ while reversing $\sigma^{yyy}$. $\ell$-counting alone would not
suffice, since it permits an even-order path through $(+3)+(-3)=0$. The fingerprint of the
mechanism is that the even harmonics appear only at $O(\gamma_3)$, carrying the $\ell{=}6{=}3{+}3$ signature of
the $w{=}\pm3$ coupling (Table~\ref{tab:ell}). Incidentally, a purely hexagonal $\ell{=}\pm6$ warping would not release at first
order; it never reaches $\ell{=}0$ from the $\{\pm1,\pm3\}$ base.

\begin{table}[t]
\caption{Angular-harmonic content of the gap-edge integrand $I(q,\theta)=\sum_\ell I_\ell e^{i\ell\theta}$ for
the AB bilayer and ABC trilayer ($\checkmark$ marks a nonzero harmonic). At $\gamma_3{=}0$ only the odd
harmonics $\ell{=}\pm1,\pm3$ are present and the scalar $\ell{=}0$ vanishes; a single action of $\gamma_3$ shifts
each harmonic by $\pm3$, populating the even harmonics and generating $\ell{=}0$ at first order, with the
$\ell{=}6{=}3{+}3$ signature of the trigonal hopping.}
\label{tab:ell}
\begin{ruledtabular}
\begin{tabular}{llcccccc}
& & $\ell{=}0$ & $\ell{=}1$ & $\ell{=}2$ & $\ell{=}3$ & $\ell{=}4$ & $\ell{=}6$\\
\hline
AB  & $\gamma_3{=}0$ & $0$ & $\checkmark$ & $0$ & $\checkmark$ & $0$ & $0$\\
    & $O(\gamma_3)$  & $\checkmark$ & $0$ & $\checkmark$ & $0$ & $\checkmark$ & $\checkmark$\\
ABC & $\gamma_3{=}0$ & $0$ & $\checkmark$ & $0$ & $\checkmark$ & $0$ & $0$\\
    & $O(\gamma_3)$  & $\checkmark$ & $0$ & $\checkmark$ & $0$ & $\checkmark$ & $\checkmark$\\
\end{tabular}
\end{ruledtabular}
\end{table}

\subsection{The emergent symmetry is approximate: the lattice residual}
\label{sec:residual}

The $\Cinf$ symmetry is exact only for the truncated $k\cdot p$ Hamiltonian, a single winding term $(\pi^\dagger)^n$ plus the diagonal mass; the full lattice does not share it, so the $\ell{=}0$ prohibition holds only within that truncation. The leading lattice correction, at the next order in $q$, is itself a trigonal ($\ell{=}\pm3$) term: it opens no new channel; it only renormalizes the $\gamma_3$ one. The residual is small and directly measurable: on the full tight-binding (TB) Hamiltonians with the skew hopping switched off ($\gamma_3{=}0$), the coherence factor is $C\approx+0.02$ for the bilayer and $+0.003$ for the trilayer [Fig.~\ref{fig:Cg3}(a)], and the kagome retains a nonzero direct channel (Secs.~\ref{sec:ab}, \ref{sec:kagome}). %This residual is carried by the lattice trigonal warping beyond the truncation, and its smallness is an independent confirmation that the forbidding symmetry is emergent and approximate.

% #####################################################################

% #####################################################################
\section{The graphene family: monolayer, bilayer, trilayer}
\label{sec:family}
% #####################################################################
We now realize the selection rule across the graphene family, where the winding number equals the number
of layers, $n{=}1,2,3$, and the releasing coupling is in every case the lattice's trigonal
warping~\cite{McCannKoshino2013,KoshinoMcCann2009}. The three systems share a single mechanism but differ in
how the band-edge weight scales with the gap. In the monolayer the response saturates;
in the bilayer and trilayer it diverges as $1/\delta$.

\emph{Model Hamiltonians.} All three stacks are TB models built from graphene's
structure factor $f(\kk)=e^{ik_ya/\sqrt3}\big[1+2\,e^{-i\sqrt3\,k_ya/2}\cos(k_xa/2)\big]$, which vanishes at the
BZ corners $K,K'$. In the layer$\,\otimes\,$sublattice basis they are block-tridiagonal, with the
$2\times2$ blocks
\begin{equation}
D_i=\begin{pmatrix}U_i & -\gamma_0 f\\ -\gamma_0 f^{*} & U_i\end{pmatrix},\qquad
V=\begin{pmatrix}\gamma_4 f & -\gamma_3 f^{*}\\ \gamma_1 & \gamma_4 f\end{pmatrix},
\label{eq:blocks}
\end{equation}
where $D_i$ is the intralayer block of layer $i$ (on-site potential $U_i$, hopping $\gamma_0$), and $V$ couples
adjacent layers (dimer $\gamma_1$, skew trigonal warping $\gamma_3$, and $\gamma_4$); a remote block
$W=\big(\begin{smallmatrix}0 & \gamma_2/2\\[1pt] 0 & 0\end{smallmatrix}\big)$ (the $\gamma_2$ hopping between
outer layers) completes the trilayer. The bilayer and trilayer are then
\begin{equation}
H_{\rm AB}=\begin{pmatrix}D_1 & V\\ V^\dagger & D_2\end{pmatrix},\qquad
H_{\rm ABC}=\begin{pmatrix}D_1 & V & W\\ V^\dagger & D_2 & V\\ W^\dagger & V^\dagger & D_3\end{pmatrix},
\label{eq:Hlattice}
\end{equation}
in the bases $(A_1,B_1,A_2,B_2)$ and $(A_1,B_1,A_2,B_2,A_3,B_3)$, with the inversion-breaking gap set by the
layer potentials $U_i$ ($\pm\delta$ for the bilayer, $(+U,0,-U)$ for the
trilayer)~\cite{McCannKoshino2013,KoshinoMcCann2009,ZhangMacDonald2010}. Near $K$,
$f\to-\tfrac{\sqrt3 a}{2}\pi^{*}$ gives the Dirac velocity $v=\tfrac{\sqrt3}{2}a\gamma_0$ and the skew velocity
$v_3=\tfrac{\sqrt3}{2}a\gamma_3$, and the eigenvector windings are $\ell=(0,1,1,2)$ and $(0,1,1,2,2,3)$.

\subsection{Monolayer (\texorpdfstring{$n=1$}{n=1}).}
\label{sec:mono}
In graphene the minimal inversion breaker that opens a gap is a
sublattice (Semenoff) mass $\delta\sigma_z$~\cite{Semenoff1984}. The gapped Dirac
cone is a winding-$1$ touching with sublattice charges $\ell=(0,1)$, so it carries the emergent $\Cinf$ and the
rank-three tensor has no $\ell{=}0$ part. The monolayer is special in two ways. First, it has no spectator band, so the response
lives entirely in the direct channel. Second, its shift vector vanishes \emph{pointwise},
$R^a=\partial_a\arg r-(A^a_{cc}-A^a_{vv})=0$ everywhere, because the dipole phase gradient identically cancels the
Berry-connection difference. At zero warping the gap-edge integrand is therefore not merely odd on average but
zero at every $\kk$.

Near $K$ the nearest-neighbor off-diagonal element expands as $-t\,f(\kk)\to v\pi^{*}+\lambda\pi^2$, giving the
gapped two-band Hamiltonian
\begin{equation}
H_{\rm mono}=\begin{pmatrix}\delta & v\pi^{*}+\lambda\pi^{2}\\[2pt] v\pi+\bar\lambda\,\pi^{*2} & -\delta\end{pmatrix},
\qquad \pi=q_x+iq_y,
\label{eq:Hmono}
\end{equation}
with $v=\tfrac32 ta$ the Dirac velocity and $\lambda=\tfrac38 ta^2$ (up to a fixed phase), both set by the same
nearest-neighbor hopping $t$. The linear $v\pi^{*}$ is the Dirac cone. The quadratic $\lambda\pi^2$, the second-order term of the expansion, is the monolayer's own trigonal
warping. Through the cross term with the Dirac cone it distorts the equal-energy contours as
$\cos 3\theta$, and it carries gauge-invariant angular charge $\ell{=}\pm3$. It is therefore the
releasing term. At $\lambda{=}0$ the integrand vanishes pointwise, as noted above, so there is no pre-existing $\ell{=}\pm3$ harmonic for the warping to convert; a single action of the warping instead generates the $\ell{=}0$ scalar directly, at first order in $\lambda$ [Eq.~\eqref{eq:mono}].

With no spectator band, the diagonal integrand collapses to the two-band form
\begin{equation}
I^{yyy}=-\frac{1}{(2E)^2}\,\Imag\!\big[\,\overline{V^y_{+-}}\,W^{yy}_{+-}\,\big],
\label{eq:monoderiv}
\end{equation}
the velocity-difference term dropping because $\Delta^y=V^y_{++}-V^y_{--}$ is real. The only $\lambda$-dependent matrix element is
the curvature $W^{yy}=\partial_{q_y}^2 H=-2\lambda\sigma_x$ (constant, already $O(\lambda)$), so $V^y=v\sigma_y$
and $\sigma_x$ are evaluated at $\lambda{=}0$. With the eigenstates of $\mathbf d\cdot\bm\sigma$
($\cos\alpha=\delta/E$) the Levi-Civita projection
$\Imag[\langle-|\sigma_a|+\rangle\langle+|\sigma_b|-\rangle]=\varepsilon_{abc}\hat d_c$ picks the mass axis,
$\Imag[\overline{V^y_{+-}}(\sigma_x)_{+-}]=v\cos\alpha=v\delta/E$, yielding
\begin{equation}
I^{yyy}(q,\theta)=\frac{\lambda\,v\,\delta}{2E^{3}},\qquad E=\sqrt{v^2q^2+\delta^2},
\label{eq:mono}
\end{equation}
which is purely $\ell{=}0$. The only $\theta$-independent piece of the pseudospin
texture is its out-of-plane component $\hat d_z=\delta/E$, tilted by the mass, and this is
the piece that the warping's interband curvature $W^{yy}$ converts into the response. The product $\lambda\delta$ makes the two
conditions manifest: the response vanishes if $\lambda{=}0$ ($\Cinf$ unbroken) or if $\delta{=}0$ (inversion unbroken). This winding-$1$ case, a warping term switching on the SC of a gapped Dirac cone
linearly in its strength, was found for the topological-insulator surface by Kim, Morimoto and
Nagaosa~\cite{KimMorimoto2017}. 
%The selection rule subsumes it as the $n{=}1$ member.

The distinctive feature is the gap scaling. We characterize it by the band-edge weight
$\mathcal{W}\equiv\int\!\frac{d^2q}{(2\pi)^2}\,I^{yyy}$, the frequency-integrated band-edge SC~\cite{Cook2017}. Although locally $I^{yyy}(q\to0)=\lambda v/(2\delta^2)$ carries the same $1/\delta^2$
as the bilayer, the explicit mass factor $\cos\alpha=\delta/E$ cancels one power in $\mathcal{W}$,
\begin{equation}
\frac{d\mathcal{W}}{d\lambda}=\frac{1}{4\pi v}\Big(1-\frac{\delta}{\sqrt{v^2Q^2+\delta^2}}\Big)
\xrightarrow[\delta\to0]{}\frac{1}{4\pi v},
\label{eq:monosat}
\end{equation}
so $\mathcal{W}$ \emph{saturates} (scaling exponent $0$) as $\delta\to0$, the two-dimensional Dirac result of
Yoshida and Murakami~\cite{YoshidaMurakami2025} [Fig.~\ref{fig:family}(a)].

\subsection{AB bilayer (\texorpdfstring{$n=2$}{n=2}).}
\label{sec:ab}
The AB bilayer has a quadratic ($n{=}2$) touching, described by the four-band Hamiltonian
$H_{\rm AB}$ [Eq.~\eqref{eq:Hlattice}] with eigenvector windings $\ell=(0,1,1,2)$. The non-dimer ends $A_1,B_2$ differ by $2$,
so the trigonal ($\ell{=}\pm3$) skew hopping $\gamma_3$ is the releasing hopping; the optical pair is the
low-energy bands $\pm E_-$ and the spectators are the dimer bands $\pm E_+$, with detuning $D_p=-E_p/E_-$
[Fig.~\ref{fig:mechanism}(b)]. The bilayer is in fact the monolayer's dual. The two carry the \emph{same} pair of windings $\{1,2\}$,
with the roles of touching and warping interchanged (App.~\ref{app:models}). Unlike in the monolayer, the
band edge is here carried by the virtual channel, the three-band Bargmann triangles of
Eq.~\eqref{eq:closed} running through the dimer spectators. At $\gamma_3{=}0$ it is near-canceled [Fig.~\ref{fig:family}(b)].

\subsubsection{Momentum-space behavior near \texorpdfstring{$K$}{K}}
\label{sec:ab_micro}
Downfolding the high-energy dimer sites gives the effective two-band Hamiltonian in the non-dimer basis
$(A_1,B_2)$,
\begin{equation}
H_{\rm eff}(\qq)=-\frac{1}{2m}\begin{pmatrix}0&(\pi^\dagger)^2\\ \pi^2&0\end{pmatrix}
+v_3\begin{pmatrix}0&\pi\\ \pi^\dagger&0\end{pmatrix}+\Delta\,\sigma_z,
\label{eq:Heff}
\end{equation}
with $\pi=q_x+iq_y$, effective mass $m\propto\gamma_1$ and trigonal velocity $v_3\propto\gamma_3$.
The two matrix elements wind by $-2$ and $+1$, so their relative winding equals the angular
charge $w{=}-3$ of Sec.~\ref{sec:weight}, now carried by the momentum factors alone. Equation~\eqref{eq:Heff}
fixes the winding bookkeeping only; the response itself is computed in the linearized
four-band model [Eq.~\eqref{eq:HAB}, App.~\ref{app:models}], which keeps the dimer
bands as explicit spectators: the virtual channel is a three-band object and has no
counterpart in the two-band form. 

\begin{figure}[t]
\centering
\includegraphics[width=0.98\linewidth]{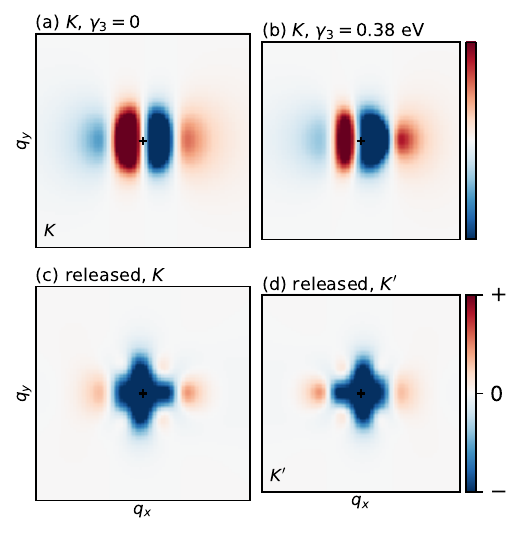}
\caption{Gap-edge SC integrand $I^{yyy}(\qq)$ around the valleys for the linearized AB bilayer. (a) At $\gamma_3{=}0$, four lobes of alternating sign whose net response vanishes, $\langle I\rangle=0$, the emergent-$\Cinf$ symmetry cancellation. (b) With $\gamma_3$ the release is a small distortion of a large, near-cancelling background. (c) Their
difference, the $O(\gamma_3)$ released component, carries a nonzero net scalar concentrated near $K$. (d) The same released component at the time-reversed valley
$K'$ [$H_{K'}(\qq)=H_K(-\qq)^{*}$]: the $\qq\!\to\!-\qq$ image of (c); the two valleys add rather than cancel, consistent with
the mirror-even $\sigma^{yyy}$.}
\label{fig:QCbilayer}
\end{figure}

Resolved in momentum space, the gap-edge integrand $I^{yyy}(\kk)$ shows four
lobes of alternating sign around $K$. At $\gamma_3{=}0$ they are mirror-symmetric about $y$ and carry only the odd harmonics $\ell{=}\pm1,\pm3$, so their net contribution vanishes; including $\gamma_3$ adds the even-harmonic distortion of Table~\ref{tab:ell}, whose
$\ell{=}0$ part, isolated in the difference maps of Fig.~\ref{fig:QCbilayer}(c,d), is the
released response concentrated near $K$.

\subsubsection{The growth is phase coherence, not amplitude}
\label{sec:ab_coherence}
\begin{figure}[t]
\centering
\includegraphics[width=0.86\columnwidth]{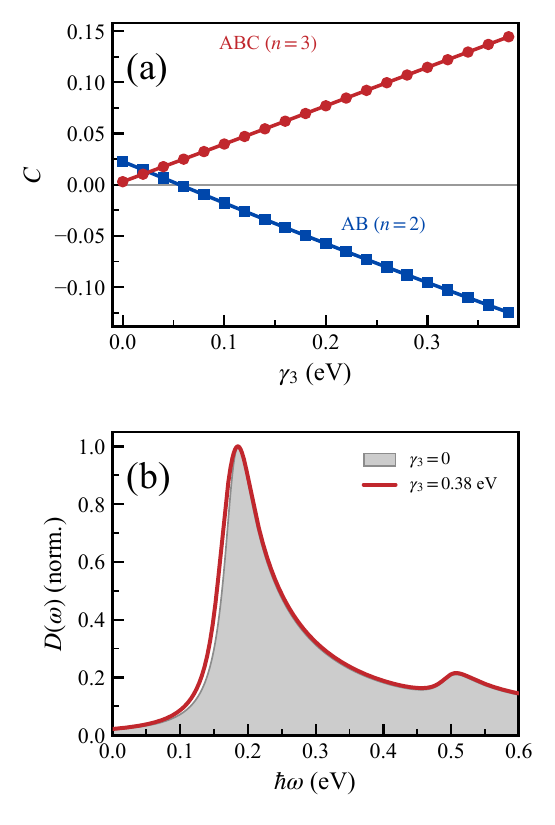}
\caption{(a) Coherence factor $C$ [Eq.~\eqref{eq:coherence}] of the band-edge SC versus trigonal
warping $\gamma_3$, from the full TB models at fixed bias
(AB, $\delta=0.10$~eV, blue; ABC, $2U=0.10$~eV, red; $n_k=1500$). Both grow linearly from a
small, symmetry-enforced residual at $\gamma_3{=}0$, but with \emph{opposite} sign [Fig.~\ref{fig:family}(d)].
(b) Dipole-only weight $D(\omega)$ of the bilayer [Eq.~\eqref{eq:dipole}, the optical rate without the shift
vector] at $\gamma_3=0$ (grey) and $\gamma_3=0.38$~eV (red), $\delta=0.10$~eV, $20$~meV
broadening.}
\label{fig:Cg3}
\end{figure}

The near-cancellation and its release are cleanly captured by the coherence factor of
Eq.~\eqref{eq:coherence}. The incoherent companion $\Avir$, the total multiband amplitude irrespective of
sign, is essentially independent of the warping, changing by only $\sim1\%$ across the scanned range
of $\gamma_3$. The signed weight $\sum_\kk\Svir$, and with it $C$, instead grows \emph{linearly} in
$\gamma_3$ from the small, symmetry-enforced residual at $\gamma_3{=}0$ [Fig.~\ref{fig:Cg3}(a)]. The
band-edge response is therefore a pure phase-coherence effect. The warping aligns the signs of the same
three-band triangles without changing their strength.

The optical absorption tells the same story. Factoring the integrand as
$I^{yyy}_{nm}=|r^y_{nm}|^2R^{y;y}_{nm}$, the dipole-only weight
\begin{equation}
D(\omega)=\frac{\pi e^3}{\hbar^2}\int_{\rm BZ}f_{nm}\,|r^y_{nm}|^2\,\delta(\omega_{nm}-\omega)\,d^2k,
\label{eq:dipole}
\end{equation}
proportional to the linear optical absorption, is likewise essentially unchanged by $\gamma_3$ (its
frequency-integrated weight varies by only $\sim4\%$, Fig.~\ref{fig:Cg3}(b)) while $\sigma^{yyy}$ grows more than fivefold. The explicit slope is derived in
Sec.~\ref{sec:closedform}.

\subsection{ABC trilayer (\texorpdfstring{$n=3$}{n=3}).}
\label{sec:abc}
The rhombohedral (ABC) trilayer realizes the same virtual mechanism on a cubic ($n{=}3$) touching. Its
eigenvector windings are $\ell=(0,1,1,2,2,3)$. The optical pair sits on the non-dimer ends $A_1$ ($\ell{=}0$)
and $B_3$ ($\ell{=}3$), which differ by the cubic-touching winding $3$, while the four dimer-derived states are
pushed to high energy by $\gamma_1$ and act as spectators. Here $A_1$ and $B_3$ are not directly bonded; their
cubic touching $(\pi^\dagger)^3$ is an effective coupling built by the chiral chain through the two dimers,
$A_1\!\to\!B_1\!\to\!A_2\!\to\!B_2\!\to\!A_3\!\to\!B_3$. The band edge is thus again carried by the
three-band Bargmann triangles. The releasing $\gamma_3$ skew hoppings act on the adjacent-layer bonds
$A_1B_2$ and $A_2B_3$, each of gauge-invariant angular charge $\pm3$ (App.~\ref{app:lcount}); a single such
hopping releases the band edge strictly linearly (Table~\ref{tab:ell}). The displacement field opens the gap ($2U$) and
$\gamma_3$ provides the warping.

\begin{figure}[t]
\centering
\includegraphics[width=0.82\linewidth]{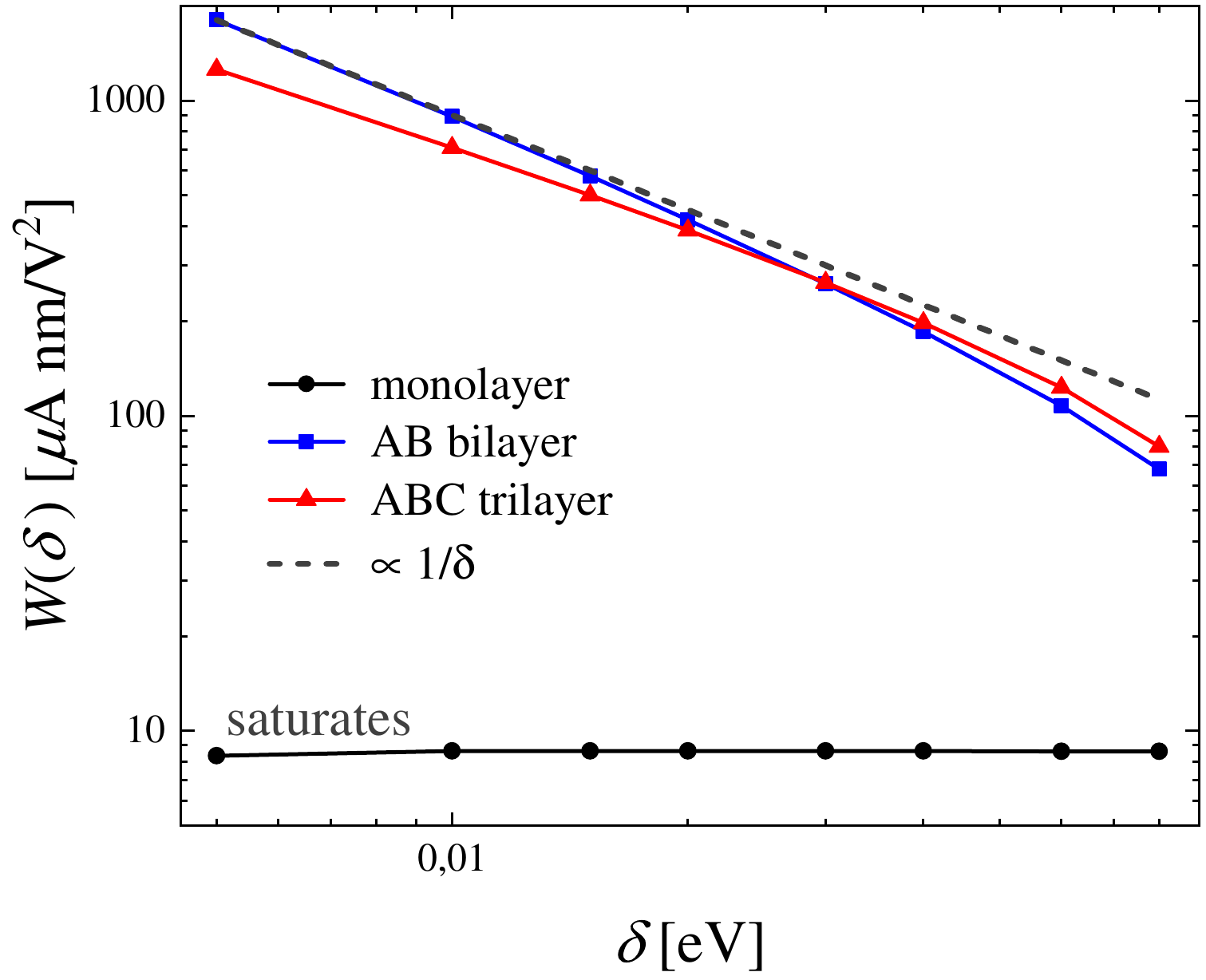}
\caption{Band-edge weight $\mathcal{W}(\delta)=\int d^2q\,I^{yyy}$ versus gap $\delta$
(log--log), from the TB spectra. The monolayer saturates (exponent $0$)
[Eq.~\eqref{eq:monosat}], while the AB bilayer and ABC trilayer both diverge as $1/\delta$
(dashed guide; trilayer fit exponents $-0.92$ to $-1.06$), the same order despite the flatter cubic touching.}
\label{fig:scaling}
\end{figure}

The band-edge weight again diverges as $1/\delta$, the same order as the
bilayer despite the flatter cubic touching, because the larger band-edge density of states is offset by a
smaller per-state three-point weight [Fig.~\ref{fig:scaling}]; the trilayer
exponent is regime-dependent, steepening toward $\delta^{-2}$ in a larger-gap,
Lifshitz-crossing window.

\begin{figure*}[t]
\centering
\includegraphics[width=0.82\textwidth]{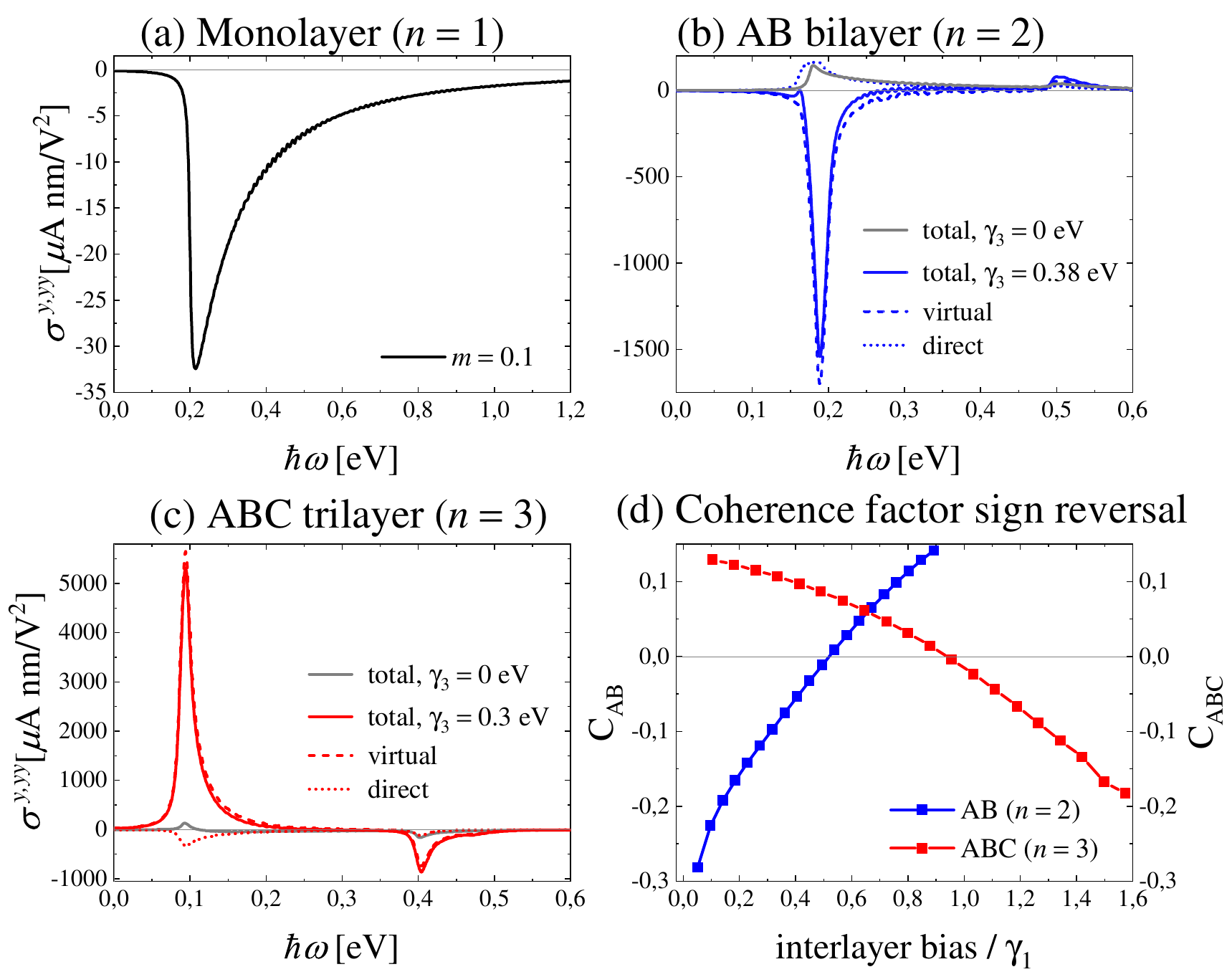}
\caption{SC across the graphene family from the full TB Hamiltonians, and its gate-driven
sign reversal. (a) Monolayer ($n{=}1$): a single direct-channel peak just above the threshold $2\delta$
($\delta=0.10$~eV). (b) AB bilayer ($n{=}2$) and (c) ABC trilayer ($n{=}3$): $\sigma^{yyy}(\omega)$ resolved into
total, virtual (dotted) and direct (dashed) channels at the physical $\gamma_3$ (color), with the total at
$\gamma_3{=}0$ (grey) for reference; the virtual channel carries essentially the whole band-edge response,
near-canceled at $\gamma_3{=}0$, and the two touchings release with \emph{opposite} sign (AB negative, ABC
positive at the band edge). (d) Coherence factor $C$ versus interlayer bias for the bilayer ($\delta/\gamma_1$,
left axis) and trilayer ($2U/\gamma_1$, right axis): $C$ crosses zero, and with it the band-edge SC
reverses sign, in opposite senses for $n{=}2$ and $n{=}3$, while the metric-like magnitude $\Avir$ stays
single-signed.}
\label{fig:family}
\end{figure*}

The trilayer band edge has, moreover, the
\emph{opposite} sign to the bilayer at small bias [Fig.~\ref{fig:family}(b,c)]. The released band-edge invariant is linear in the effective
chiral coupling of the $N$-layer rhombohedral stack, which alternates as $(-1)^{N-1}v^N/\gamma_1^{N-1}$. Each
downfolding of a dimer pair brings a factor $-1/\gamma_1$, so the effective touching is
$-v^2/\gamma_1\,(\pi^\dagger)^2$ for the bilayer and $+v^3/\gamma_1^2\,(\pi^\dagger)^3$ for the
trilayer~\cite{KoshinoMcCann2009,ZhangMacDonald2010}. The same alternation fixes the touching's $N\pi$
Berry phase, and it is directly visible in the opposite slopes of the coherence factor
$C(\gamma_3)$ for the two stacks [Fig.~\ref{fig:Cg3}(a)].

Rhombohedral trilayers provide an immediate test. A large SC that reverses sign under the
displacement field has been reported~\cite{ABCtrilayer2024}. This is precisely the behavior of the
signed invariant. As in the bilayer, the band-edge response reverses when the field drives $C$
through zero [Fig.~\ref{fig:family}(d)].

% #####################################################################
\section{Analyzing the band-edge shift current in the AB bilayer}
\label{sec:closedform}
% #####################################################################
Sections~\ref{sec:selection} and~\ref{sec:family} established the selection rule and
followed it through the graphene family. We now evaluate the bilayer release by a hybrid symbolic--numerical scheme that treats the
angular and radial variables separately. Because each eigenvector carries a definite phase
$e^{i\ell_j\theta}$, the angular average is a single Fourier coefficient (the $z^{0}$ term,
$z=e^{i\theta}$), which we extract symbolically and without approximation; the remaining radial
integral over $u=v|q|$ is then evaluated numerically. The selection rule and the band-edge
integrand are thereby exact, while the frequency-integrated slope reduces to one numerical
radial quadrature. We use the linearized four-band AB model. Its two
dimer bands are explicit spectators, and a single first-order perturbation theory in the
warping delivers the band-edge integrand, its slope, and the origin of the gate-tunable sign
reversal.

\subsection{First-order perturbation theory}
\label{sec:pt}
At $\gamma_3=0$ the linearized bilayer bands are angle independent,
\begin{equation}
E_\pm(u)=\Big[\delta^2+\tfrac{\gamma_1^2}{2}+u^2
\pm\sqrt{\tfrac{\gamma_1^4}{4}+\gamma_1^2u^2+4\delta^2u^2}\,\Big]^{1/2},
\label{eq:bands}
\end{equation}
with $u=v|q|$ the dimensionless radial momentum measured from $K$. The optical pair are
the low-energy bands $m,n=\mp E_-$ (direct gap $2E_-$); the spectators that build the
virtual channel are the dimer bands $p=\pm E_+$, for which the detuning weight of
Eq.~\eqref{eq:closed} collapses to $D_p=-E_p/E_-$, a real function of $q$ alone. The four eigenvectors carry definite angular charges
$\ell=(0,1,1,2)$ under the emergent $\Cinf$, so the angular average of any quantity is
its $z^{0}$ Fourier coefficient ($z\equiv e^{i\theta}$); we write
$I_0(u)\equiv z^{0}[\,\cdot\,]$ for the resulting band-edge integrand.

Section~\ref{sec:vanish} showed that $\Svir$ vanishes identically at $\gamma_3=0$. We turn on
the skew hopping $H_{A_1B_2}=v_3\pi$ as a first-order perturbation and evaluate the band-edge
integrand by computer algebra (\texttt{SymPy}~\cite{SymPy2017}). We verify that the zeroth-order term cancels, and the
surviving $O(\gamma_3)$ term is an $\ell=3\!\to\!0$ conversion, so the release is
\emph{strictly linear} in $\gamma_3$. 

The band-edge integrand $I_0(u)=z^{0}[\partial_{\gamma_3}\Svir]$
is a rational function of the band energies $E_\pm(u)$; it is too lengthy to display here. Its small-gap limit, however, is compact,
\begin{equation}
I_0(u\!\to\!0;\delta)=-\frac{v^{3}}{2\gamma_1\delta^{2}},
\label{eq:I0}
\end{equation}
negative for every $\delta$ and diverging as $1/\delta^{2}$ as the gap closes. Equation~\eqref{eq:I0}
is exact, and the numerical calculations reproduce it.

\subsection{The \texorpdfstring{$1/\delta$}{1/delta} divergence of the bilayer slope}
\label{sec:slope}
To convert the band-edge integrand into the measured initial slope
$\sigma^{yyy}\simeq B(\delta)\,\gamma_3$ we integrate $I_0(u)$ radially. In the
small-gap limit $\delta\ll\gamma_1$, the rescaling $u=\sqrt{\delta\gamma_1}\,x$ collapses
the band-edge region onto a $\delta$-independent shape $F(x)$, so the slope there is set by
the single number $\mathcal{J}=\int_0^\infty x\,F(x)\,dx$; the integral is not elementary
and evaluates numerically to $\mathcal{J}\approx0.39$, giving
\begin{equation}
B(\delta)\simeq-\frac{\sqrt3\,a}{4\pi}\,\frac{\mathcal{J}}{\delta}\qquad(\delta\ll\gamma_1).
\label{eq:Bslope}
\end{equation}
Equation~\eqref{eq:Bslope} is the small-gap asymptote of $B$, not its full bias
dependence. At larger $\delta$ the finite-momentum region left out of the collapse competes
with the band edge and eventually reverses the sign of $B$ (Sec.~\ref{sec:reversal}). 

%The slope is independent of the Dirac velocity, and hence of $\gamma_0$, for a structural reason. The intralayer and skew hoppings enter the lattice model through the same structure factor $f(\kk)$ [Eq.~\eqref{eq:blocks}], so their velocities share the geometric prefactor, $v=\tfrac{\sqrt3}{2}a\gamma_0$ and $v_3=\tfrac{\sqrt3}{2}a\gamma_3$, and the warping enters the linearized model only through the ratio $\varepsilon=v_3/v=\gamma_3/\gamma_0$. The three dipoles of the Bargmann triangle supply the factor $v^3$ of the band-edge integrand, and it cancels against the $v^{-3}$ assembled by the momentum measure ($q\,dq=u\,du/v^{2}$) and by the chain rule $\partial_{\gamma_3}=\tfrac{\sqrt3\,a}{2v}\,\partial_\varepsilon$; of the velocity factors, only the geometric length $\tfrac{\sqrt3}{2}a$ survives in $B$. The cancellation requires $v_3$ and $v$ to share the same prefactor; a warping term detached from the Dirac velocity would not show it.
The $1/\delta$ divergence is the generic two-dimensional signature of the
quadratic touching: a band edge that would saturate to a constant for a linear cone
($n=1$) instead diverges as the inverse gap for the $n=2$ touching, the
quadratic-band-touching enhancement identified by Yoshida and
Murakami~\cite{YoshidaMurakami2025}. The radial integral reproduces the full-lattice slope (App.~\ref{app:pt}).

\subsection{Gate-tunable sign reversal and which momenta carry it}
\label{sec:reversal}
We now show that the sign of the band-edge photocurrent is gate-tunable. The
mechanism is traced in detail for the bilayer through the band-edge integrand of
Sec.~\ref{sec:pt} (exact in the angle, numerical in the radius) and confirmed on the full
TB models of both the bilayer and the trilayer [Fig.~\ref{fig:family}(d)].

\begin{figure}[t]
\centering
\includegraphics[width=0.9\linewidth]{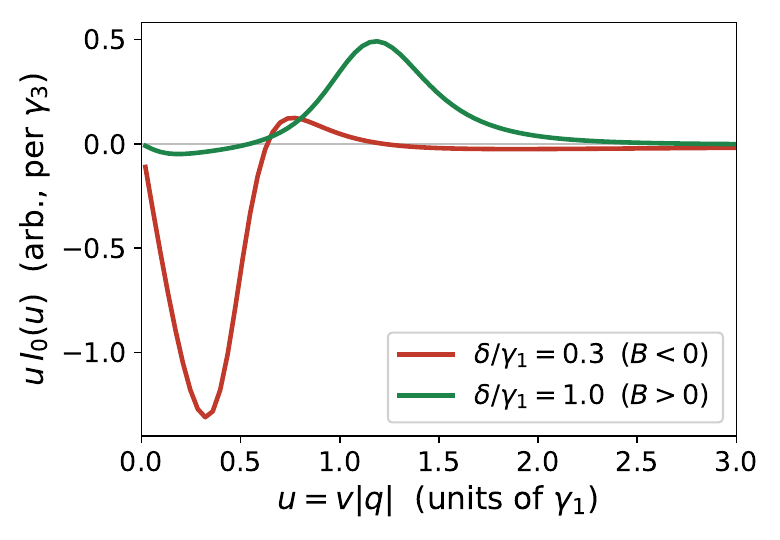}
\caption{Radial profile of the per-valley band-edge integrand $u\,I_0(u)$ for the
linearized AB bilayer (first order in $\gamma_3$, $v=\gamma_1=1$). A negative band-edge
lobe ($u\to0$) competes with a positive lobe near $u\sim\gamma_1$. For
$\delta/\gamma_1=0.3$ the negative lobe dominates ($B<0$); for $\delta/\gamma_1=1.0$ the
positive lobe dominates ($B>0$); the radial integral vanishes at
$\delta/\gamma_1\approx0.65$, the gate-tunable sign reversal.}
\label{fig:signflip}
\end{figure}

In the bilayer the reversal comes with a subtlety. The band-edge integrand [Eq.~\eqref{eq:I0}] is negative for \emph{all} $\delta$, so the reversal is not a sign change of the band edge itself; it lives in the radial integral. The radial profile
$u\,I_0(u)$ (Fig.~\ref{fig:signflip}) is negative at the band edge ($u\to0$, the
$1/\delta^{2}$ lobe) but turns positive near $u\sim\gamma_1$, and the frequency-integrated
weight $B$ is set by the competition between the two lobes. At small gap the sharp negative
band-edge lobe wins ($B<0$); as $\delta\to\gamma_1$ it flattens and the positive
finite-momentum lobe takes over ($B>0$). Within the linearized, first-order-in-$\gamma_3$
band-edge functional the two balance at $\delta/\gamma_1\approx0.65$.

A complementary, fully nonperturbative diagnostic is the whole-zone coherence factor
$C$ [Eq.~\eqref{eq:coherence}], evaluated on the full TB model with $\gamma_3 \neq 0$
(Fig.~\ref{fig:family}). It too changes sign with bias, crossing zero at
$\delta/\gamma_1\approx0.52$. The two crossings are close but distinct; $B$ is a band-edge, $O(\gamma_3)$,
frequency-integrated quantity, whereas $C$ is obtained by integrating over the entire BZ. 

In both descriptions, the incoherent magnitude $\Avir$ never vanishes across the
reversal, so the entire sign change resides in the signed three-point invariant. %A positive, metric-based figure of merit cannot reproduce it, a point we take up in Sec.~\ref{sec:metric}.

The trilayer behaves the same way. The whole-zone coherence factor also crosses zero as the displacement field $2U$ is swept [Fig.~\ref{fig:family}(d)], and the band-edge current reverses in the \emph{opposite} sense, the $n{=}2$ versus $n{=}3$ contrast of Sec.~\ref{sec:abc}.

% #####################################################################
\section{The kagome lattice}
\label{sec:kagome}
% #####################################################################
The graphene family releases the band-edge current by breaking the emergent $\Cinf$ with a
real-space warping term. The kagome lattice reaches the same result by a different route.
Its flat band breaks the emergent symmetry \emph{intrinsically}, taking over the role that
the $\gamma_3$ hopping plays in the multilayers. The three-point invariant of
Eq.~\eqref{eq:closed} therefore governs this band edge as well, now with the flat band
itself as the spectator of the Bargmann triangle. Since the gap-edge touching has winding
$n{=}1$, of the monolayer class, the kagome carries both mechanisms of the family on a
single touching. The direct channel is supplied automatically by the trigonal anisotropy
that the lattice hoppings carry beyond leading order in $q$, the counterpart of the
monolayer's $\lambda\pi^2$. The virtual channel runs through the flat band. No warping term is added by hand in our kagome calculations. The
continuum model is truncated at leading order and its direct channel vanishes accordingly;
the full-lattice calculation contains the anisotropy automatically
(Table~\ref{tab:kagome}). The $\lambda\pi^2$ reference is an analogy of role only. An
inversion-broken kagome SC has recently been reported~\cite{Liu2025kagomeSC}.

\subsection{Model and inversion breaking}
\label{sec:kag_model}
The nearest-neighbor kagome TB model (sublattices $A,B,C$; Bravais vectors
$a_1=(1,0)$, $a_2=(\tfrac12,\tfrac{\sqrt3}{2})$) has off-diagonal blocks
\begin{equation}
h_{AB}=-\big(t_\triangle\, e^{i\kk\cdot a_1/2}+t_\triangledown\, e^{-i\kk\cdot a_1/2}\big)
\quad\text{(and cyclic)}.
\label{eq:kag}
\end{equation}
One band remains perfectly flat at $E=t_{\triangle}+t_{\triangledown}$ for every breathing ratio. The hexagonal loop states interfere destructively triangle by triangle, and the interference survives the breathing because each triangle contains only one of the two hoppings. At the isotropic point $t_{\triangle}=t_{\triangledown}=t$ the spectrum also shows Dirac cones at $K$ and a quadratic touching of the flat and dispersive bands at $\Gamma$ [Fig.~\ref{fig:kagome}(a)].

Inversion acts differently here
than in the honeycomb. The kagome inversion centers sit at the hexagon centers and map
each sublattice to itself ($A\!\to\!A$, $B\!\to\!B$, $C\!\to\!C$), so a sublattice onsite
pattern $(\varepsilon_A,\varepsilon_B,\varepsilon_C)$ \emph{preserves} inversion and gives
zero SC. The minimal inversion breaker is instead the breathing distortion
$t_\triangle\neq t_\triangledown$, which exchanges the up- and down-triangles related by
that inversion. It gaps the $K$ Dirac (turning on $\sigma^{yyy}$) and leaves the $\Gamma$
touching gapless. The gap-edge pair is thus a winding-$1$ Dirac of the monolayer
class, two bands whose sublattice charges $\ell$ differ by one, with a $\pi$ Berry phase in
the gapless limit.

\subsection{The flat band as an intrinsic, self-releasing spectator}
\label{sec:kag_release}
What is new is the third band. The flat band is the \emph{sole spectator} of the gap-edge
pair. It carries the remaining $C_3$ charge and lies \emph{outside} the Dirac pair's
winding ladder. Monolayer graphene has no such spectator, so the kagome is the one family
member with a winding-$1$ Dirac \emph{and} a virtual channel.

The cleanest way to see this is at zeroth order in the warping. As in the monolayer, the
two-band shift vector of the gap-edge Dirac vanishes pointwise across the whole BZ ($R^{a}=0$ everywhere), so the direct channel is zero in the continuum and is
switched on only by the lattice trigonal warping, the same $\ell=\pm3$ release as the
monolayer $\lambda\pi^{2}$. Unlike in the monolayer, however, the integrand does not vanish;
the multiband triangle through the flat band survives. This is precisely the virtual
channel of Eq.~\eqref{eq:closed}, the bilayer/ABC mechanism, but with a flat band as the
spectator.

The distinction between the two routes is fixed by which bands share the emergent isotropy. The $\Cinf$ is never a symmetry of the crystal. It is an accidental isotropy of the leading $k\cdot p$ Hamiltonian, and its selection rule operates only as long as every band entering the response shares it.  In the graphene multilayers the whole relevant multiband block is
isotropic at leading order, and the crystal's trigonal ($C_3$) structure enters only as a
higher-order correction, the $\ell=\pm3$ warping $\gamma_3$; there is therefore a
$\Cinf$ window at the band edge in which the $\ell=0$ scalar is forbidden. The gapped Dirac
pair of the kagome inherits the same accidental isotropy \emph{as a two-band subspace}, hence the
pointwise vanishing noted above. 

The flat band, however, injects the crystallographic $C_3$ of the little group at $K$ already at leading order. At $K$ the three bands carry the three distinct $C_3$ charges, defined only mod $3$, and the two dipoles that link the flat band to the Dirac pair close the Bargmann triangle with total angular charge $\pm3$. Under the would-be $\Cinf$ this is $\ell=\pm3$, not a scalar, and its angular average would vanish; under the actual $C_3$ it is $0\bmod3$, a lattice scalar. The three-band model is therefore only $C_{3v}$-symmetric, never $\Cinf$-symmetric, and there is no isotropic window. The flat-band triangle carries the $\ell{=}0$ response already in the continuum, with no warping term.

Table~\ref{tab:kagome} puts numbers to this. In the continuum the direct channel is identically zero while the virtual, flat-band channel is finite. In the full-lattice calculation, whose hoppings carry the trigonal anisotropy automatically, the direct channel appears with equal weight.

\begin{table}[h]
\caption{Band-edge integrand $I_0$ of the breathing kagome, split into direct and virtual
channels. In the linearized continuum (gapped $K$ Dirac plus flat band, no trigonal anisotropy)
the direct channel vanishes identically while the flat-band triangle already carries the
$\ell=0$ scalar; in the full lattice the direct channel appears with equal weight.}
\centering
\begin{ruledtabular}
\begin{tabular}{lcc}
 & direct $I_0$ & virtual (flat band) $I_0$\\
\hline
continuum model       & $0$      & $+0.073$\\
full lattice   & $+0.070$ & $+0.070$\\
\end{tabular}
\end{ruledtabular}

\label{tab:kagome}
\end{table}

\begin{figure}[t]
\centering
\includegraphics[width=\linewidth]{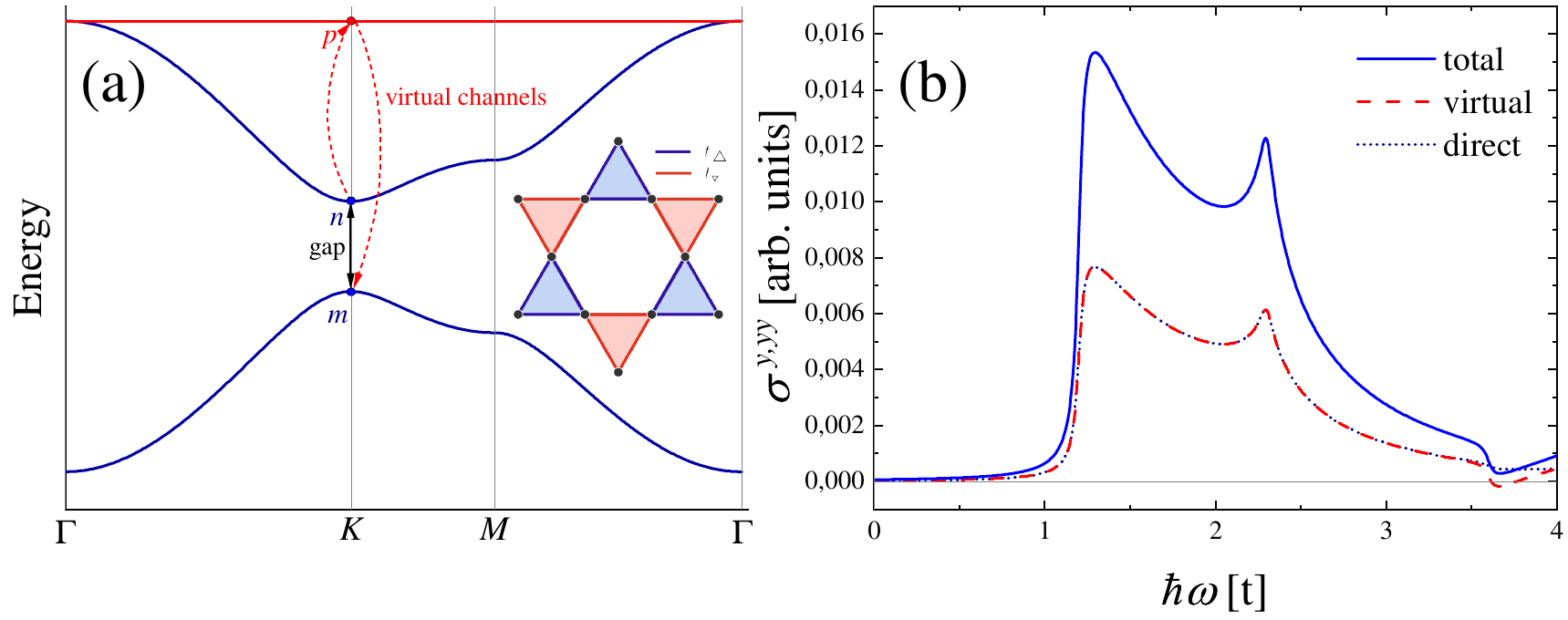}
\caption{Kagome flat-band complement. (a) Schematic bands: the breathing
$t_{\triangle}\neq t_{\triangledown}$ gaps the $K$ Dirac (optical pair $m,n$) and leaves the
flat band at $t_{\triangle}+t_{\triangledown}$ as the sole spectator (dashed loop $m\!\to\!\mathrm{flat}\!\to\!n$);
inset, the breathing lattice. (b) SC $\sigma^{yyy}(\omega)$ resolved into total,
virtual and direct channels; the two are comparable, locked to one half each at
the $K$-dominated threshold.}
\label{fig:kagome}
\end{figure}

The direct and virtual channels nearly coincide across the threshold
[Fig.~\ref{fig:kagome}(b)]. A structural identity underlies this near-equality. The
band-edge response
is dominated by a small disk about $K$ (over $90\%$ of the threshold weight), and \emph{at}
$K$ the two channels are strictly equal, each carrying one half of the integrand. This is a
nearest-neighbor, $C_3$-protected identity, exact for every breathing ratio
$t_\triangle/t_\triangledown$; we derive it in App.~\ref{app:kaglock}. Away from $K$ the
equality relaxes, so the frequency-integrated weights differ at the few-percent level
(integrated ratio $\approx0.93$ at $t_\triangle/t_\triangledown=1.5/0.5$), which is why we
call the two channels comparable rather than identical.

% #####################################################################
\section{Assessing the quantum metric as a figure of merit}
\label{sec:metric}
% #####################################################################
Maximizing the QMT has been proposed as a design strategy for a large
SC~\cite{Chen2024,TormaPeottaBernevig2022}. Inter-orbital mixing opens virtual channels,
the Wannier functions delocalize, and the metric grows. We show that, for the
Dirac-like family, the band-edge response is controlled by $\Svir$, not by the QMT. This
section quantifies that statement on the three members of the family.

\subsection{Two-band magnitude versus three-band sign}
\label{sec:two_vs_three}
The two quantities differ in the number of Bloch states they involve. The QMT
correlates two states at a time, $\Svir$ three, and only the second structure
can retain a gauge-invariant phase. The QMT and its BZ-integrated trace are built from
\emph{pairs} of dipoles,
\begin{equation}
g^{ab}_n=\Real\!\sum_{m\neq n} r^a_{nm}\,r^b_{mn},\qquad
\Omega_I=\!\!\sum_{n\in\mathrm{val}}\!\int_{\rm BZ}\!\!\frac{d^2k}{(2\pi)^2}\,
\mathrm{Tr}\,g_n ,
\label{eq:QGT}
\end{equation}
and the trace is a sum of squared dipole moduli, hence positive. In the multiband
members studied here, the band-edge response is instead carried by $\Svir$ [Eq.~\eqref{eq:closed}], a triple product of dipoles weighted by the
sign-indefinite detuning $D_p$ (Secs.~\ref{sec:family} and~\ref{sec:kagome}). A three-point
correlator carries signed terms. It can be small through cancellation even when every dipole
is large, and it can reverse under a parameter sweep. A two-point correlator of squared
moduli admits no cancellation; it is small only if every dipole is small, and it never
changes sign. The same first-order correction $v^{(1)}$ enters the two through different
structures. The metric receives $|v^{(1)}|^2$, which accumulates without cancellation; the
current receives the cross term $\Real[v^{(0)*}v^{(1)}]$, which carries a sign. Hence the
metric scales quadratically in $\gamma_3$ while the current scales linearly. This
difference in power law shows up twice. At small warping the linear term dominates, so the band-edge current
switches on faster than the metric, which stays flat near $\gamma_3{=}0$. And a quadratic
dependence is even in $\gamma_3$, so the metric cannot reproduce the oddness
$\sigma^{yyy}(-\gamma_3)=-\sigma^{yyy}(\gamma_3)$ of Sec.~\ref{sec:selection}. Reversing
$\gamma_3$ is not an experimental knob, but it makes explicit why a positive, even
quantity cannot serve as a figure of merit for a signed response.

\subsection{Three tests on the Dirac-like family}
\label{sec:falsify}
\begin{figure}[t]
\centering
\includegraphics[width=\linewidth]{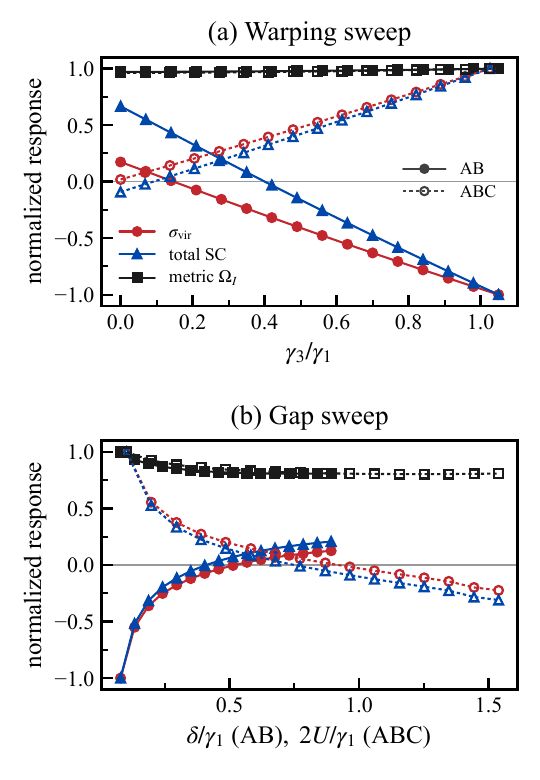}
\caption{Behavior of $\Svir$ and $\Omega_I$ for the AB (solid
lines, filled symbols) and the ABC (dashed lines, open symbols), from the full
TB Hamiltonians. Each panel overlays $\Svir$ (red circles), the total SC (blue triangles), and the valence $\Omega_I$ of Eq.~\eqref{eq:QGT} (dark squares), each normalized to its
own maximum modulus. (a) Warping sweep $\gamma_3/\gamma_1$ at fixed gap ($\delta=0.10$~eV for
AB, $2U=0.10$~eV for ABC). (b) Gap sweep at the physical warping ($\gamma_3=0.38$~eV for AB,
$0.30$~eV for ABC). The sweep variable is the interlayer bias $\delta/\gamma_1$ for AB
and the displacement field $2U/\gamma_1$ for ABC. $\Svir$ and the total current cross zero at nearby but not identical
values, the small offset set by the smooth direct-channel background.}
\label{fig:metric}
\end{figure}

The family provides three independent comparisons (Table~\ref{tab:metric},
Fig.~\ref{fig:metric}).

In the \emph{AB bilayer}, sweeping $\gamma_3$ at fixed gap drives $\Svir$ linearly
through zero and into sign reversal while the valence metric $\Omega_I$ rises monotonically
[Fig.~\ref{fig:metric}(a)]. If the metric set the response, the ratio
$\sigma^{\rm vir}/\Omega_I$ would remain constant; instead it varies by an order of magnitude
across the scan. The four-band manifold is
fully connected already at $\gamma_3=0$, so the metric accumulates $|v^{(1)}|^2$ while the
current follows the cross term.

In the \emph{ABC trilayer}, sweeping the displacement field reverses the sign of
$\Svir$ (zero crossing at $2U/\gamma_1\approx0.92$) while $\Omega_I$ remains positive and
smooth across the reversal [Fig.~\ref{fig:metric}(b)]. The complementary scans, the trilayer
in $\gamma_3$ [Fig.~\ref{fig:metric}(a)] and the bilayer in bias [Fig.~\ref{fig:metric}(b)],
show the same pattern.

In the \emph{kagome} the contrast is sharpest, and symmetry enforces it. The same knob
that moves the metric leaves the response untouched. The hexagon-centered inversion maps each
sublattice to itself, so any on-site pattern $(\varepsilon_A,\varepsilon_B,\varepsilon_C)$
preserves inversion, while the breathing $t_\triangle\neq t_\triangledown$ is the
perturbation that breaks it. Sweeping the traceless pattern $(2\varepsilon,-\varepsilon,-\varepsilon)$
with $\varepsilon$ up to $0.4t$ at fixed $t_\triangle=t_\triangledown$ therefore keeps the
inversion-odd $\Svir$ at zero at every step, while $\Omega_I$
varies by more than an order of magnitude and peaks at the band touching that the on-site
pattern induces near $\varepsilon\approx0.05t$. The sign structure that decides the band edge is
not contained in $\Omega_I$.

\begin{table}[t]
\caption{The three tests of Sec.~\ref{sec:falsify}: the knob varied in each system,
and the resulting behavior of the band-edge invariant $\Svir$ against $\Omega_I$. For the kagome, the on-site potentials $\varepsilon_j$ are swept, keeping $t_\triangle=t_\triangledown$. In every case $\Omega_I$ stays positive
and featureless where $\Svir$ is linear, sign-reversing, or held at zero by symmetry.}
\label{tab:metric}
\begin{ruledtabular}
\begin{tabular}{llll}
system & knob & $\Svir$ & $\Omega_I$ \\
\hline
AB     & $\gamma_3$ & linear, reverses & positive, monotone \\
ABC    & $2U$       & reverses         & positive, smooth \\
kagome & on-site $\varepsilon_j$ & $0$ by inversion & nonzero, varies \\
\end{tabular}
\end{ruledtabular}
\end{table}

% #####################################################################
\section{Experimental proposal}
\label{sec:experiment}
% #####################################################################
The bilayer SC sign reversal translates into a concrete, gate-controlled experiment. In dual-gated
Bernal bilayer graphene, the two gates act independently. The total gate voltage holds the
sample at charge neutrality while the displacement field $D$ opens the interlayer gap
$2\delta$ without doping. The reversal window $\delta/\gamma_1\approx0.5$--$0.65$ of
Sec.~\ref{sec:reversal} corresponds, for $\gamma_1\approx0.38$~eV, to an interlayer bias
$2\delta\approx0.4$--$0.5$~eV, within reach of hBN-encapsulated devices.

At fixed photon energy near threshold ($\hbar\omega\gtrsim2\delta$, in
the mid-infrared) and for linear polarization, sweeping $D$ drives the band-edge
photocurrent $\sigma^{yyy}$ through zero. The reversal is a threshold in $\delta/\gamma_1$
rather than a smooth gate dependence, and the mirror $M_x$ confines it to the $\sigma^{yyy}$
component, separating it from injection
and photothermal backgrounds. There have been several previous proposals along these lines~\cite{Zheng2023, CentrosymBPVE2025,Fregoso2019}. 

The same measurement extends to other structures. In the ABC trilayer the displacement
field tunes the gap through $2U$ (Sec.~\ref{sec:reversal})~\cite{ABCtrilayer2024,GiantSC2025};
the inversion-broken kagome metals, in which an SC has recently been
measured~\cite{Liu2025kagomeSC}, are a further platform. In each case the geometry is the
same: normal-incidence linearly polarized light, a DC short-circuit photocurrent along the
direction fixed by the mirror, and intensity linearity confirming the SC origin.

% #####################################################################
\section{Discussion and outlook}
\label{sec:discussion}
% #####################################################################
A single selection rule explains the four cases. At a gapped winding-$n$ Dirac-like touching in a $C_3$-symmetric two-dimensional crystal, an emergent continuous rotation $\Cinf$ forbids the band-edge SC; trigonal
warping breaks that symmetry and releases the current linearly, through the unique coupling
of gauge-invariant angular charge $w=\pm3$. The released response is quantified by the signed, detuning-weighted three-point Bargmann
invariant $\Svir$, whose bounded companion $C$ is the band-edge figure of merit. The monolayer releases through its two-band direct channel
alone; the AB bilayer and ABC trilayer through their dimer spectators; the breathing kagome
through its flat band, which supplies the crystallographic $C_3$ in place of the warping
hopping. The gate-driven sign reversal, in opposite directions for the two stacks, and the
small-gap $1/\delta$ divergence are shared by the bilayer and the trilayer; for the bilayer
the law is moreover explicit and velocity-independent [Eq.~\eqref{eq:Bslope}].

The results are demonstrated within one symmetry class, gapped Dirac and
flat-band touchings in two dimensions with a $C_3$ little group; establishing $\Svir$
as a figure of merit beyond this class would require a broader survey. Magnitude bounds
are complementary to it. The Tan--Rappe upper limit on
$|\sigma^{\rm shift}|$~\cite{TanRappe2019} constrains how large the response can be,
and no upper bound anticipates the zero crossing that the gate drives.

Although the theory rests on a single-particle Hamiltonian and a clean, defect-free
crystal, so that disorder, excitonic, and many-body corrections will dress the threshold and
may shift the reversal locus, the central result stands. The band-edge SC of a gapped
two-dimensional Dirac-like system is set by a symmetry selection rule and a geometric invariant.
% #####################################################################
\appendix
% #####################################################################

\section{Model Hamiltonians, bases, and windings}
\label{app:models}
This appendix records the explicit linearized $k\cdot p$ Hamiltonians and derives the
eigenvector windings quoted in Sec.~\ref{sec:family}; the full lattice forms are
Eqs.~\eqref{eq:blocks}--\eqref{eq:Hlattice} and the monolayer is Eq.~\eqref{eq:Hmono}. In the
near-$K$ expansion of Sec.~\ref{sec:family} ($f\to-\tfrac{\sqrt3a}{2}\pi^{*}$,
$\pi=q_x+iq_y$), the bilayer reads in the basis $(A_1,B_1,A_2,B_2)$, dropping $\gamma_4$, whose charge $w{=}0$ cannot release the band edge,
\begin{equation}
H_{\rm AB}=\begin{pmatrix}-\delta & v\pi^* & 0 & v_3\pi\\ v\pi & -\delta & \gamma_1 & 0\\
0 & \gamma_1 & +\delta & v\pi^*\\ v_3\pi^* & 0 & v\pi & +\delta\end{pmatrix},
\label{eq:HAB}
\end{equation}
and the trilayer is the block-tridiagonal assembly of Eq.~\eqref{eq:Hlattice} with the
near-$K$ blocks (dropping $\gamma_2,\gamma_4$)
\begin{equation}
D_i=\begin{pmatrix}U_i & v\pi^*\\ v\pi & U_i\end{pmatrix},\qquad
V=\begin{pmatrix}0 & v_3\pi\\ \gamma_1 & 0\end{pmatrix},
\label{eq:kpblocks}
\end{equation}
with layer potentials $(U_1,U_2,U_3)=(+U,0,-U)$, in the basis $(A_1,B_1,A_2,B_2,A_3,B_3)$.

The windings follow from the emergent-symmetry condition [Eq.~\eqref{eq:emergent}]. Each
component carries $e^{i\ell_j\theta}$ with $\ell_j$ fixed by requiring $H_0$ (at
$\gamma_3{=}0$) to commute with $J=-i\partial_\theta-\hat L$. The intralayer $v\pi^{*}\sim
e^{-i\theta}$ links sites whose charges differ by one, and each dimer $\gamma_1$ links equal
charges, which fixes
\begin{equation}
\ell_{\rm AB}=(0,1,1,2),\qquad \ell_{\rm ABC}=(0,1,1,2,2,3),
\label{eq:windings}
\end{equation}
so the non-dimer ends $A_1,B_N$ differ by the touching winding $n$. For the breathing
kagome the little group at $K$ is $C_3$ and the three sublattices carry the distinct charges
$\{2,1,0\}\bmod 3$; the flat band is the spectator. The gauge-invariant angular charges of the
warping couplings are verified in App.~\ref{app:lcount}.

\section{\texorpdfstring{$\ell$}{l}-counting of the warping couplings and Fourier checks}
\label{app:lcount}
Section~\ref{sec:weight} states the master rule $w=(\ell_a-\ell_b)-w_{\rm hop}$; here we
evaluate it for each warping coupling and record the numerical checks. In the bilayer and
trilayer the skew hopping $\gamma_3$ couples $A_1\!\leftrightarrow\!B_2$ (and
$A_2\!\leftrightarrow\!B_3$) with momentum factor $\pi$ ($w_{\rm hop}{=}{+}1$). From the
windings of Eq.~\eqref{eq:windings}, $w=(\ell_{A_1}-\ell_{B_2})-1=(0-2)-1=-3$. In the monolayer
the releasing coupling is the second-order term $\lambda\pi^{2}$ of the expansion, linking
$A\!\leftrightarrow\!B$ ($\ell=0,1$) with $w_{\rm hop}{=}{+}2$, giving
$w=(0-1)-2=-3$ as well. Both realize the unique $w{=}\pm3$ channel. A direct
rotation-and-gauge test of Eq.~\eqref{eq:weight} returns $w=-3$ to numerical precision.
By contrast $\gamma_4$ couples sublattices already connected by $\gamma_0,\gamma_1$ and
carries $w{=}0$, so it does not release the band edge (App.~\ref{app:numerics}).

The angular decomposition of the gap-edge integrand confirms the mechanism numerically. At
$\gamma_3{=}0$ the scalar vanishes, $|I_{\ell=0}|\sim10^{-15}$, while the odd harmonics are
$O(1)$ ($|I_{\ell=1}|=18$ and $|I_{\ell=3}|=6.3$ for the ABC gap-edge pair). Switching on
$\gamma_3$ populates, at $O(\gamma_3)$, just the harmonics obtained by shifting this odd
base by $\pm3$, each traceable to its parent: $\ell=0$ from $3-3$, $\ell=2$ from $-1+3$,
$\ell=4$ from $1+3$, and $\ell=6$ from $3+3$, with no other lines appearing
(Table~\ref{tab:ell}). The $\ell=6$ line has a single possible parentage, the base harmonic $\ell=3$ shifted
up by $+3$, so its appearance at first order certifies a coupling that transfers three
units of angular momentum, no more and no fewer. Finally, the
full conductivity integrand carries an extra optical factor $|r^b|^2$. At $\gamma_3{=}0$
this is a rotational scalar and leaves the $\ell$-grading of the bare rank-three tensor
unchanged, so the counting of Sec.~\ref{sec:vanish} applies to the measured response as
well.

\section{First-order perturbation theory and symbolic evaluation}
\label{app:pt}
This appendix records the first-order theory sketched in Sec.~\ref{sec:pt}. We treat the skew
hopping as the perturbation, $H_1=\partial H/\partial\varepsilon$ with $\varepsilon\equiv v_3/v$,
whose only nonzero element is $H_{A_1B_2}=v_3\pi=\varepsilon\,u\,z$. First order in $\varepsilon$
dresses both levels and states, with level shift $E^{(1)}_j=\langle j|H_1|j\rangle$ and state
shift $|\delta j\rangle=\sum_{k\neq j}\langle k|H_1|j\rangle/(E^{(0)}_j-E^{(0)}_k)\,|k\rangle$.
Because $H_1$ carries winding $\pm1$ and the eigenvectors carry $e^{i\ell_j\theta}$, the diagonal
level shift is stamped with the trigonal harmonic, $E^{(1)}_j\propto z^{\pm3}$.

The interband dipoles inherit the dressing. With $r_{IJ}=v^y_{IJ}/(i\,\omega_{IJ})$, $\omega_{IJ}=E_I-E_J$,
the expansion $r_{IJ}=r^{(0)}_{IJ}+\varepsilon\,r^{(1)}_{IJ}$ has
\begin{equation}
r^{(1)}_{IJ}=\frac{v^{y(1)}_{IJ}}{i\,\omega^{(0)}_{IJ}}
-\frac{v^{y(0)}_{IJ}\,\omega^{(1)}_{IJ}}{i\,\big(\omega^{(0)}_{IJ}\big)^{2}},
\label{eq:r1}
\end{equation}
with $\omega^{(1)}=E^{(1)}_I-E^{(1)}_J$ and $v^{y(1)}_{IJ}$ collecting the corrected-bra,
corrected-ket, and $\partial_yH_1$ contributions. Inserting $D_p=D^{(0)}+\varepsilon D^{(1)}$ and
$T_p=r_{mn}r_{np}r_{pm}=T_0+\varepsilon T_1$ into $\Svir=\sum_p D_p\,\Real T_p$ gives the
$O(\varepsilon)$ coefficient of the band-edge integrand,
\begin{equation}
\Svir^{(1)}=\!\!\sum_{p=\pm E_+}\!\!\big[D^{(1)}\Real T_0+D^{(0)}\Real T_1\big],
\label{eq:assembly}
\end{equation}
and $I_0(u)=z^{0}[\Svir^{(1)}]$. The evaluation is symbolic (\texttt{SymPy}~\cite{SymPy2017}). The phases
$e^{i\ell_j\theta}$ are kept as powers of $z=e^{i\theta}$, so the angular average is the exact
$z^{0}$ coefficient, while the radial variable $u$ and the parameters $(\delta,\gamma_1,v)$ are
evaluated numerically. The zeroth-order piece drops from the average,
$z^{0}[\sum_p D^{(0)}\Real T_0]=0$, because $T_0$ carries only $\ell=\pm1,\pm3$; this is the
$\Cinf$ cancellation of Sec.~\ref{sec:vanish} made explicit. The surviving $\ell=3\!\to\!0$
term has two opposite-sign pieces of the same order, $D^{(1)}\Real T_0$ and $D^{(0)}\Real T_1$ (at
$u=0.6$, $+46.6$ and $-132.6$), so the release is carried neither by the detuning nor by the
dipole dressing alone. The resulting $I_0(u)$ is a rational function of $E_\pm(u)$ that does not
reduce to a compact form. Its $u\to0$ limit, Eq.~\eqref{eq:I0}, follows exactly in four
steps.

\emph{(i) Eigensystem at $u=0$.} With no momentum,
$H_0=\mathrm{diag}(-\delta,-\delta,\delta,\delta)$ plus the dimer $\gamma_1$ on the
$B_1$--$A_2$ block. The optical pair is pure, $m=A_1$ (energy $-\delta$) and $n=B_2$
($+\delta$), with gap $2\delta$; the spectators are the eigenstates of the $2\times2$
dimer block $\gamma_1\sigma_x-\delta\sigma_z$, with energies $E_{p_\pm}=\pm\Omega$,
$\Omega=\sqrt{\gamma_1^2+\delta^2}$, and components on $(B_1,A_2)$
\begin{equation}
|p_+\rangle=\frac{(\gamma_1,\ \Omega+\delta)}{\sqrt{2\Omega(\Omega+\delta)}},\qquad
|p_-\rangle=\frac{(\gamma_1,\ -(\Omega-\delta))}{\sqrt{2\Omega(\Omega-\delta)}}.
\label{eq:pspec}
\end{equation}

\emph{(ii) The assembly collapses to a single term.} The perturbation matrix is itself
proportional to $u$ ($H_{A_1B_2}=\varepsilon\,u\,z$), so at $u=0$ every dressing
vanishes identically, $E^{(1)}_j=0$, $|\delta j\rangle=0$, $\omega^{(1)}=0$,
$D^{(1)}=0$. Of the first-order dipole only the explicit $\partial_yH_1$ term survives,
$r^{(1)}_{IJ}=\langle I|\partial_yH_1|J\rangle/(i\,\omega^{(0)}_{IJ})$, nonzero only
on the optical bond $A_1$--$B_2$. At the same time $r^{(0)}_{mn}=0$, because
$\partial_yH_0$ has no $A_1$--$B_2$ element; hence $T_0=0$, the last two terms of $T_1$
vanish with it, and Eq.~\eqref{eq:assembly} reduces to
\begin{equation}
\Svir^{(1)}\big|_{u=0}=\sum_{p=p_\pm}D^{(0)}_p\,\Real\big[r^{(1)}_{mn}\,r^{(0)}_{np}\,r^{(0)}_{pm}\big].
\label{eq:collapse}
\end{equation}

\emph{(iii) The three factors, exactly.} The vertex dipole is
$r^{(1)}_{mn}=\langle A_1|\partial_yH_1|B_2\rangle/(i\,\omega_{mn})
=iv/[\,i(-2\delta)]=-v/(2\delta)$. For the virtual legs, $B_2$ couples only to the
$A_2$ component of the spectator and $A_1$ only to its $B_1$ component (both with matrix
element $iv$), so, with $\omega_{np_\pm}=\delta\mp\Omega$ and
$\omega_{p_\pm m}=\delta\pm\Omega$,
\begin{equation}
r^{(0)}_{np_\pm}=\frac{\pm iv\sqrt{\tfrac{\Omega\pm\delta}{2\Omega}}}{i(\delta\mp\Omega)},\qquad
r^{(0)}_{p_\pm m}=\frac{iv\,\gamma_1/\sqrt{2\Omega(\Omega\pm\delta)}}{i(\delta\pm\Omega)},
\label{eq:legs}
\end{equation}
whose product collapses, using $\Omega^2-\delta^2=\gamma_1^2$, to
$r^{(0)}_{np_\pm}r^{(0)}_{p_\pm m}=\mp v^2/(2\Omega\gamma_1)$. The detuning is exact
at $u=0$, $D^{(0)}_{p_\pm}=(\delta-\delta\mp2\Omega)/(2\delta)=\mp\Omega/\delta$.

\emph{(iv) Sum over the two spectators.} All factors are real, and each spectator
contributes the same amount,
\begin{equation}
\Big(\mp\frac{\Omega}{\delta}\Big)\Big(-\frac{v}{2\delta}\Big)
\Big(\mp\frac{v^2}{2\Omega\gamma_1}\Big)=-\frac{v^{3}}{4\gamma_1\delta^{2}},
\label{eq:halfsum}
\end{equation}
the factor $\Omega$ cancelling, so the sum over $p_\pm$ gives Eq.~\eqref{eq:I0}
exactly, for every $\delta$ and not only for $\delta\ll\gamma_1$. Numerical
calculations confirm the limit to $5\times10^{-5}$ across $\gamma_1$, $v$, and
$\delta$.

\section{Role of \texorpdfstring{$\gamma_4$}{gamma4} in the AB bilayer}
\label{app:numerics}
Scanning $\gamma_4$ from $0$ to $0.28$~eV at fixed
$\gamma_3=0.38$~eV and $\delta=0.10$~eV changes the virtual channel by less than $0.5\%$,
against the order-of-magnitude effect of $\gamma_3$. This is the selection rule at work. The
$\gamma_4$ hopping couples sublattice sectors already connected by $\gamma_0$ and $\gamma_1$, so its
angular charge is $w{=}0$ and it modifies existing velocity matrix elements rather than
generating the $\ell{=}0$ scalar. It enters the metric-like $|v^{(1)}|^2$ but not the signed
band-edge release.

Concretely, restoring $\gamma_4$ to the linearized bilayer of Eq.~\eqref{eq:HAB} adds an
adjacent-layer, same-sublattice velocity on the $A_1A_2$ and $B_1B_2$ bonds,
\begin{equation}
H_{\rm AB}^{\gamma_4}=\begin{pmatrix}
-\delta & v\pi^* & -v_4\pi^* & v_3\pi\\[1pt]
v\pi & -\delta & \gamma_1 & -v_4\pi^*\\[1pt]
-v_4\pi & \gamma_1 & +\delta & v\pi^*\\[1pt]
v_3\pi^* & -v_4\pi & v\pi & +\delta\end{pmatrix},\qquad
v_4=\tfrac{\sqrt3}{2}a\gamma_4,
\label{eq:HABg4}
\end{equation}
the near-$K$ image of the $\gamma_4 f$ entries of the interlayer block $V$ in
Eq.~\eqref{eq:blocks} ($f\to-\tfrac{\sqrt3a}{2}\pi^{*}$). These terms carry a single power of
$\pi^{*}\!\sim e^{-i\theta}$, the same winding as the Dirac hopping
($w_{\rm hop}{=}{-}1$), and they link $\ell_{A_1}{=}0$ to $\ell_{A_2}{=}1$, giving angular
charge $w=(\ell_{A_1}-\ell_{A_2})-w_{\rm hop}=(0-1)+1=0$. With $w{=}0$, $\gamma_4$ dresses
the eigenvectors without opening an $\ell{=}3\!\to\!0$ release channel; this is why the scan
above leaves $\Svir$ essentially unchanged.

\section{Direct\,$=$\,virtual identity at the kagome \texorpdfstring{$K$}{K} point}
\label{app:kaglock}
The identity of this appendix emerged as a curiosity of the kagome analysis. We record it
for completeness; it may deserve a deeper exploration elsewhere. We establish the statement
used in Sec.~\ref{sec:kag_release}, that the direct and virtual channels of the
breathing-kagome band edge coincide at $K$ for every breathing ratio. Write the diagonal integrand as $I_0=\Imag[\,\bar r^{y}_{nm}\,r^{y}_{nm;y}\,]$ and
split the generalized derivative Eq.~\eqref{eq:gend} into its resonant (direct) and
spectator (virtual) parts, Eqs.~\eqref{eq:direct} and~\eqref{eq:virt}. At a band extremum
($\Delta_y=v^y_{nn}-v^y_{mm}=0$) the direct part reduces to the interband curvature
$W^{yy}_{nm}=\langle n|\partial_y^2H|m\rangle$ and the virtual part to the spectator sum, so
the two channels coincide if and only if
\begin{equation}
W^{yy}_{nm}=-S,\qquad
S=\!\!\sum_{p\neq n,m}\!\! v^y_{np}v^y_{pm}\Big(\tfrac1{\omega_{pm}}-\tfrac1{\omega_{np}}\Big).
\label{eq:lock}
\end{equation}
Three facts, all fixed by the $C_3$ little group at $K$, establish the identity.

\emph{(i) Representation content.} The little group at $K$ is $C_3$ (a $120^\circ$ rotation
times the sublattice cycle $A\!\to\!B\!\to\!C$); its three one-dimensional irreps assign the
band charges $\{0,1,2\}\bmod 3$. Because the velocity $v_y$ is a vector it has no
$C_3$-invariant diagonal, so $\langle n|v_y|n\rangle=0$ and $\Delta_y=0$; $K$ is a band
extremum of the optical pair.

\emph{(ii) Chiral selection.} The chiral second derivatives $\partial_\pm^2H$
($\partial_\pm=\partial_x\pm i\partial_y$) carry angular momentum $\pm2$, and at $K$ a matrix element is nonzero only if the operator's charge matches the bands' charge difference mod 3. The gapped Dirac
pair differs in charge by one unit, so only one chirality connects it,
$\langle n|\partial_-^2H|m\rangle=0$ while $\langle n|\partial_+^2H|m\rangle\neq0$. The
interband curvature is a single term, and the same selection leaves a single charge-compatible
spectator, the flat band, so the sum $S$ collapses to one triangle.

\emph{(iii) Nearest-neighbor closure.} For the pure nearest-neighbor kagome the interband
curvature and the single spectator triangle are tied by $\langle n|\partial_+^2H|m\rangle=4S$,
giving $W^{yy}_{nm}=-S$. The direct and virtual parts of $r^y_{nm;y}$ are then equal, each
$(i/\omega_{nm})S$, so each channel carries one half of the band-edge integrand.

We confirmed the identity numerically at $K$. The ratio $W^{yy}_{nm}/(-S)$ equals unity to
finite-difference precision, independent of the breathing ratio
$t_\triangle/t_\triangledown$, and $\langle n|\partial_-^2H|m\rangle=0$ as required. The lock
is thus a nearest-neighbor, symmetry-protected identity rather than fine-tuning. It is
special to the $C_3$-invariant $K$ point. A $C_3$-symmetric longer-range hopping preserves
steps (i) and (ii) but detunes the closure of step (iii) (the ratio $W^{yy}_{nm}/(-S)$ moves
to $-0.63$ and $-1.38$ for second-harmonic amplitudes $s=0.1,0.3$), and at the
$C_2$-invariant $M$ point the analogous ratio is $1/3$. Because the optical threshold is
$K$-dominated, the spectral peak inherits direct\,$=$\,virtual, while the frequency-integrated
weights differ at the few-percent level.

\section*{Data availability}
The data behind all figures of this article are openly available in the Zenodo
repository of Ref.~\cite{ZenodoSC}.

\section*{Code availability}
The tight-binding codes, the symbolic-computation scripts, and a document with the
complete analytical derivations are openly available in the same Zenodo
repository~\cite{ZenodoSC}.

\begin{acknowledgments}
L.E.F.F.T. acknowledges financial support by ANID FONDECYT (Chile) through grant 1250751.
E.S.M. acknowledges financial support from Proyecto Interno USM PI\_LIR\_26\_12.
L.C. acknowledges financial support from grant PID2022-136285NB-C31 funded
by MCIN/AEI/10.13039/501100011033/.
\end{acknowledgments}

\section*{Author contributions}
F.P.R., M.C.S., and E.S.M. performed the calculations. F.P.R. and E.S.M. wrote the
manuscript. L.C. and L.E.F.F.T. contributed to the discussion of the results and
revised the manuscript. All authors approved the final version.

\section*{Competing interests}
The authors declare no competing interests.

\section*{AI-use disclosure}
An AI assistant (Claude, Anthropic, including the Opus 4.8 and Fable 5 models, accessed
through the Claude desktop and Claude Code interfaces) was used to assist with
condensation, structuring, and language editing of the manuscript and with plotting
scripts for the figures. The assistant was directed through interactive prompting, and
every AI-suggested edit and script was reviewed and verified by the authors. All physical
content, derivations, and numerical results were produced and verified by the authors, who
take full responsibility for the final text.

\bibliography{ref}

\end{document}